\documentclass[conference,lettersize,10.5pt]{IEEEtran}
\usepackage{makecell}
\usepackage[english]{babel}
\usepackage[T1]{fontenc}
\usepackage[noadjust]{cite}
\usepackage{url,color} % Citation numbers being automatically sorted and properly "compressed/ranged".
\usepackage{graphics,amsfonts}
\usepackage{epstopdf}
\usepackage[pdftex]{graphicx}
\usepackage[cmex10]{amsmath}
\usepackage{hyperref}
\usepackage{mathtools}
\DeclarePairedDelimiter{\ceil}{\lceil}{\rceil}
\usepackage[utf8]{inputenc}
\usepackage{times}
\begin{document}
%%%%%%%%%%%%%%%%%%%%%%%%%%%%%%%%%%%%%%%%%%
\title{Complex Network Analysis of Men Single ATP Tennis Matches}

\author{\IEEEauthorblockN{Umberto Michieli}
\IEEEauthorblockA{Department of Information Engineering, University of Padova -- Via Gradenigo, 6/b, 35131 Padova, Italy\\{\tt michieli@dei.unipd.it}
}}

\maketitle

\begin{abstract}

Who are the most significant players in the history of men tennis? Is the official ATP ranking system fair in evaluating players scores? Which players deserved the most contemplation looking at their match records? Which players have never faced yet and are likely to play against in the future? Those are just some of the questions developed in this paper supported by data updated at April 2018\footnote{The datasets used in this paper are made publicly available at: \hyperref[https://drive.google.com/open?id=1mCxZfkkpIC9o-nxZ1yW3GBBdvBOPW6mQ]{https://drive.google.com/open?id=1mCxZfkkpIC9o-nxZ1yW3GBBdvBOPW6mQ} \cite{data}}.\\
In order to give an answer to the aforementioned questions, complex network science techniques have been applied to some representations of the network of men singles tennis matches. Additionally, a new predictive algorithm is proposed in order to forecast the winner of a match. 
\end{abstract}

\begin{IEEEkeywords} Tennis, Complex Network, Ranking, Link Prediction, Community Dectection.
 \end{IEEEkeywords}

%%%%%%%%%%%%%%%%%%%%%%%%%%%%%%%%%%%%%%%%%
\section{Introduction and Related Works}\label{sec:intro}
During the last decades Network Science field has been rediscovered and addressed as the "new science" \cite{barabasibook}, \cite{lewis}. A lot of issues have been (re-)examined thanks to Network Science techniques, which are nowadays permeating the way we face the world as a unique interconnected component. The presence and the immediate availability of a huge amount of digital data describing every kind of network and the way in which its nodes interact, has made possible an interdisciplinary analysis of many large-scale systems.\\
Similar techniques have been recently applied also to professional sports, in order to discover complex interactions phenomena and universal rules which are almost invisible and difficult to recognize restricting the attention to small networks or to microscopic level. For example, complex-network analysis were conducted on soccer (e.g. in \cite{soccer_premier} and \cite{soccer_brazilian}), football (\cite{football1} and \cite{football_attacking}), basket (\cite{basket} and \cite{basket_nba}), baseball (\cite{baseball}) and cricket (\cite{cricket1} and \cite{cricket2}), just to name a few.

In professional tennis as well, there are few studies examining how to map matches into complex networks and then developing new ranking methods alternative to the ATP (Association of Tennis Professionals) official one.\\
The first work of this kind is represented by \cite{situngkir}, where the authors explained the network generation and then they performed some simple analysis on single Grand Slams tournaments matches only (i.e. four tournaments each year: respectively \textit{Australian Open, Roland Garros, Wimbledon} and \textit{U.S. Open}). Then an important contribution was brought by \cite{radicchi}, where a different network modeling is proposed and the PageRank algorithm is applied identifying \textit{Jimmy Connors} as the most important single tennis player between 1968 and 2010.\\
More recent tennis-related complex network studies regard new ranking methods proposal and evaluations (see as reference \cite{subgraph}, \cite{spanias} and \cite{maquirriain}), or are related to doubles matches \cite{double} or to the gender and handedness effects in top ranking positions \cite{breznik}.

On the other side, however, in literature there is not an exhaustive and precise explanation about the network topology of the tennis matches graph. Moreover, some papers seems to be hasty in asserting a scale-free nature of the network with some inaccuracies. In this study the resulting network structure when all the official men single tennis matches are considered since the so-called "Open Era" to the end of March 2018 (i.e. from 1968 onward; the ATP organization, instead, was founded in 1972) is carefully analyzed and all its major properties are stated, which can be exploited for some interesting structural considerations, even not touched in the existing studies, and for further analysis.

In the second part of the paper some ranking algorithms have been applied aiming at confirming the present literature on ranking methods, thus seeing how active tennis players have improved their overall prestige over the recent years. However, at the same time, the aim of the paper is to provide some useful considerations about link prediction and communities detection.\\ All the computer simulations were performed in Matlab.

\section{Generation of Dataset and Network}\label{sec:dataset}

The first step is the generation of the dataset: all men tennis matches since 1968 are considered. The data can be freely downloaded directly from the ATP website \cite{ATP} and from other online repositories (like \cite{tennisco} for recent data and \cite{abstract}) allowing to fix some inconsistencies in the official ones. Hence the first step to do is to merge the small datasets, provided on a yearly basis, in one only; this is a very delicate operation since we need to account for many format differences and bring them all back to a common language for the information's specification. For some of the next considerations the following features of interest for each match have been kept: the tournament level, the tournament stage, the winner player and the loser player. \\A brief excursion follows in order to explain those quantities. The tournament levels allows to identify the importance of a match, in fact ATP hosts tournaments of very different prizes (as regards both money and ATP ranking points assigned), which in increasing order of importance are: \textit{ATP 250} tournaments, \textit{ATP 500}, \textit{Masters 1000}, the annual \textit{ATP World Tour Finals} and the \textit{Grand Slams} (different names were used in the past but similar considerations hold). The ATP points assigned to the winner of the tournament are respectively 250, 500, 1000, 1500 and 2000 and lower points are attributed to the players in proportion to the reached stage of the tournament (e.g round 128, round 64, up to semifinal and final); refer to Table \ref{tab:points} for a simplified overview of current points attribution distribution where points of qualified players are taken into account as last rounds of each entry (more detailed system points attribution can be examinated in \cite{dingle}).

\begin{figure}[]
\centering
\hspace{0cm}\includegraphics[width=0.8\linewidth]{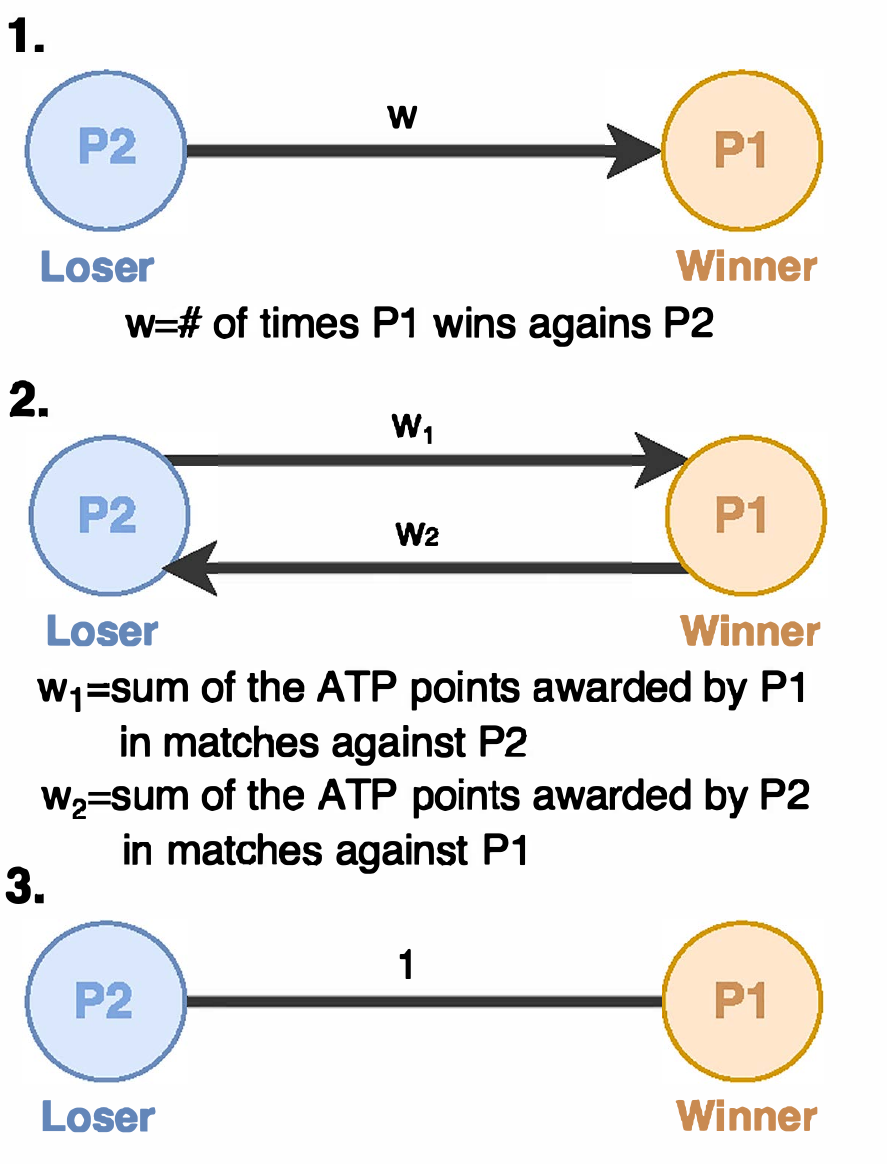}
\caption{Three different network representations considered; each edge is associated to a respective weight briefly explained.}
\label{fig:representations}
\end{figure}

\begin{table*}[]
\centering
\label{tab:points}
\begin{tabular}{c|c|c|c|c|c|c|c|c|}
\cline{2-9}
                                            & W    & F    & SF       & QF       & R16      & R32      & R64     & R128     \\ \hline
\multicolumn{1}{|l|}{Grand Slam}            & 2000 & 1200 & 720      & 360      & 180      & 90       & 45      & 10       \\ \hline
\multicolumn{1}{|l|}{ATP World Tour Finals} & +500 & +400 & \multicolumn{6}{l|}{+200 points for each round robin match win} \\ \hline
\multicolumn{1}{|l|}{Masters 1000}          & 1000 & 600  & 360      & 180      & 90       & 45       & 25      & 15       \\ \hline
\multicolumn{1}{|l|}{ATP 500}               & 500  & 300  & 180      & 90       & 45       & 20       & 10      &          \\ \hline
\multicolumn{1}{|l|}{ATP 250}               & 250  & 150  & 90       & 45       & 20       & 10       & 5       &          \\ \hline
\end{tabular}
\caption{ATP points distribution. W=Winner, F=Finalist, SF=Semi Finalist, QF=Quarter Finalist, R=Round.}
\end{table*}
With those considerations in mind it is possible to map the matches into many different network representations (in all of them, however, the nodes represent the players, hence they are homogeneous) and the main scenarios are summarized in Figure \ref{fig:representations} matching the following descriptions:
\begin{enumerate}
\item \textbf{Direct graph representation:} in this model an edge exists from every loser to the winner, each link has a weight equal to the number of times the destination node won over the starting node. In case of multiple links the weights are just summed. Similar representations were adopted in \cite{radicchi} considering data up to 2010, in \cite{situngkir} considering data between 90s and 00s of male and female matches of Grand Slams only with different weights function, and in \cite{subgraph} with data of top-100 players only and different weights function. The obtained graph is not symmetric, not even if the respective unweighted version is considered.
\item \textbf{Direct and symmetric graph representation:} this original proposal assumes the existence of a directed link between each couple of nodes which played at least one match against each other. The weights are the respective ATP points awarded by the two players; note that even the loosing player gets a non-negative points score. Also in this representation in case of multiple links the points are summed up. By construction, the network will be structurally symmetric but with possibly very different weights. 
\item \textbf{Undirect (and unweighted) graph representation:} for many considerations, however, the most useful representation is the one where two players are connected through an undirected and unweighted edge if they play at least one match against each other, thus we obtain an undirected and symmetrical network.
\item \textbf{Extended version of undirect and unweighted graph:} this is the largest possible dataset since official tennis matches have been established because it takes into account also \textit{Davis Cup}, \textit{Challenger} and \textit{Futures} tournaments since 1991, which were not considered in the first three datasets. 
Two players are connected through an undirected and unweighted edge if they play at least one match against each other.
\end{enumerate}

The first three datasets are formed by $N=4245$ nodes (i.e. players) and a total of $151734$ matches which leads to $L=170168$ or $101436$ links depending on the selected representation (larger number for the second representation). The fourth dataset comprises of $N=22405$ players and $L=998114$ links.

Notice that, as in many real networks, the matrix can still be defined as sparse since it holds $L\ll L_{max}=\displaystyle \frac{N(N-1)}{2}$ links, where $L_{max}$ is the maximum number of links of a network with N nodes.\\These large datasets will allow us to spot general trends and most competitive players overall; for more specific analysis it is enough just to restrict the attention to a smaller period of time (e.g. if we are interested in a specific player we should consider restricting our focus to his career epoch). For construction, some results can be inherently biased toward the already retired players but in practice we will see that this does not always hold because of, for example, the increasing number of tournaments and of ATP points assigned each year.

\section{Topological Results} \label{sec:res}

In this section some results of complex network techniques are explored highlighting the properties and the underlying physical meaning. Moreover some comparisons among the different network representations will be asserted to verify the common aspects through different views.

\subsection{Adjacency Matrix}

\begin{figure}[]
\hspace{-0cm}\includegraphics[width=\linewidth]{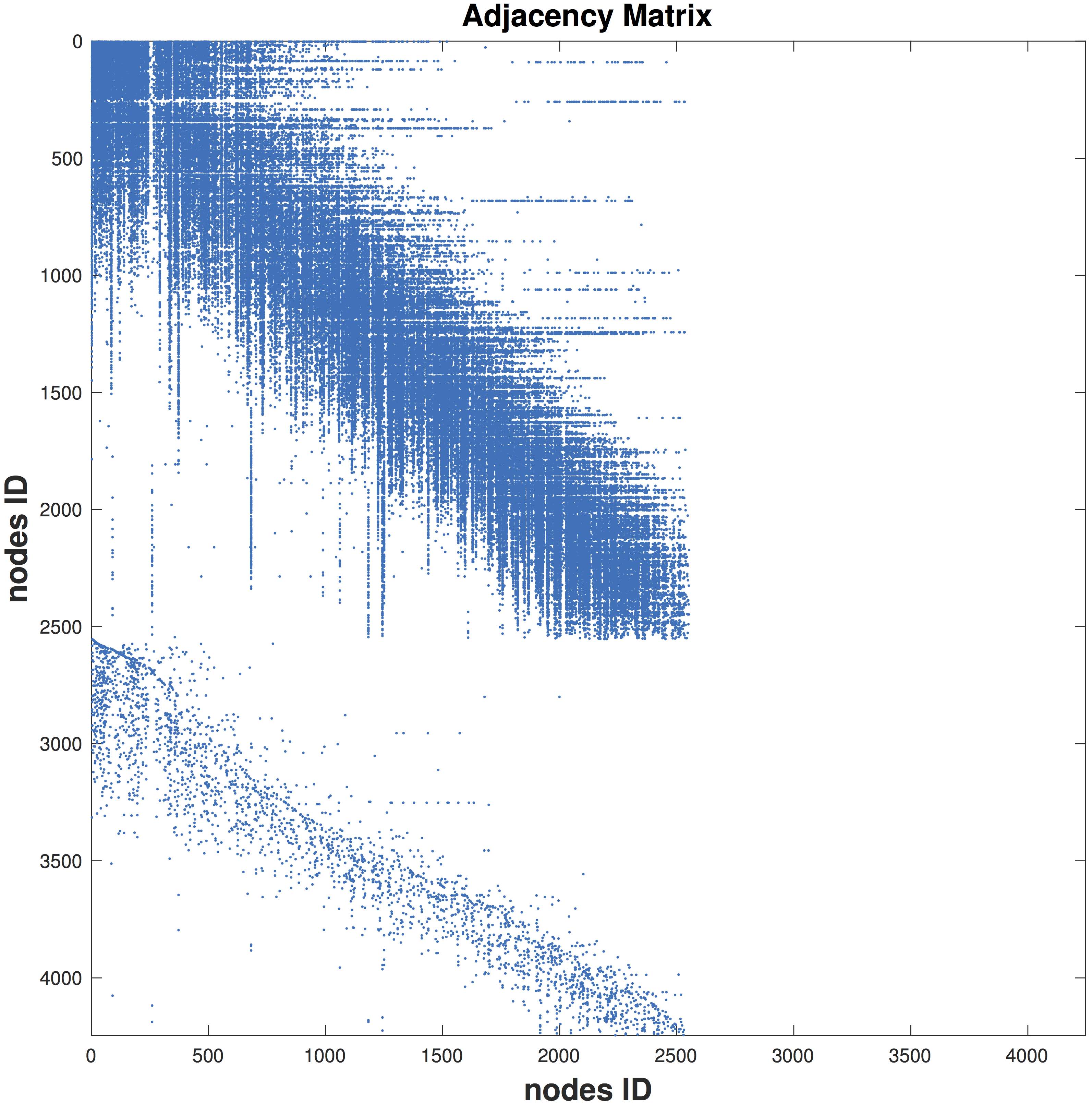}
\caption{Adjacency matrix of the direct network representation.}
\label{fig:adjacency}
\end{figure}

In Figure \ref{fig:adjacency} the adjacency matrix of the first representation of direct network is shown. 
The plot of the matrix has this shape because, by construction, first of all the players which have won at least one match are inserted and after that the players which figure only for lost matches are considered. Thus, since there can't be any link between two always-loser players, the bottom-right part of the matrix is composed by all-zero entries. 
The bottom-left part of the matrix (and the respective up-right part if symmetric case is considered) do have a few points which are the matches lost by players who only have lost matches in the higher ATP tournaments (they surely have won some matches in minor ATP tournaments like \textit{Challengers} or \textit{Futures} in order for them to be admitted in the main draw of the most important ATP tournaments).\\ Moreover notice that the columns and the rows with a lot of non-zero entries are associated with players who have faced a lot of different players, thus usually they are players with a long-career and very \textit{successful}, we should come back to this idea of evaluation of successful player in section \ref{sec:res_nature} and in \ref{sec:res_ranking}.

\subsection{Network Visualization and Small World Property}

\begin{figure}[]
\hspace{-0cm}\includegraphics[width=\linewidth]{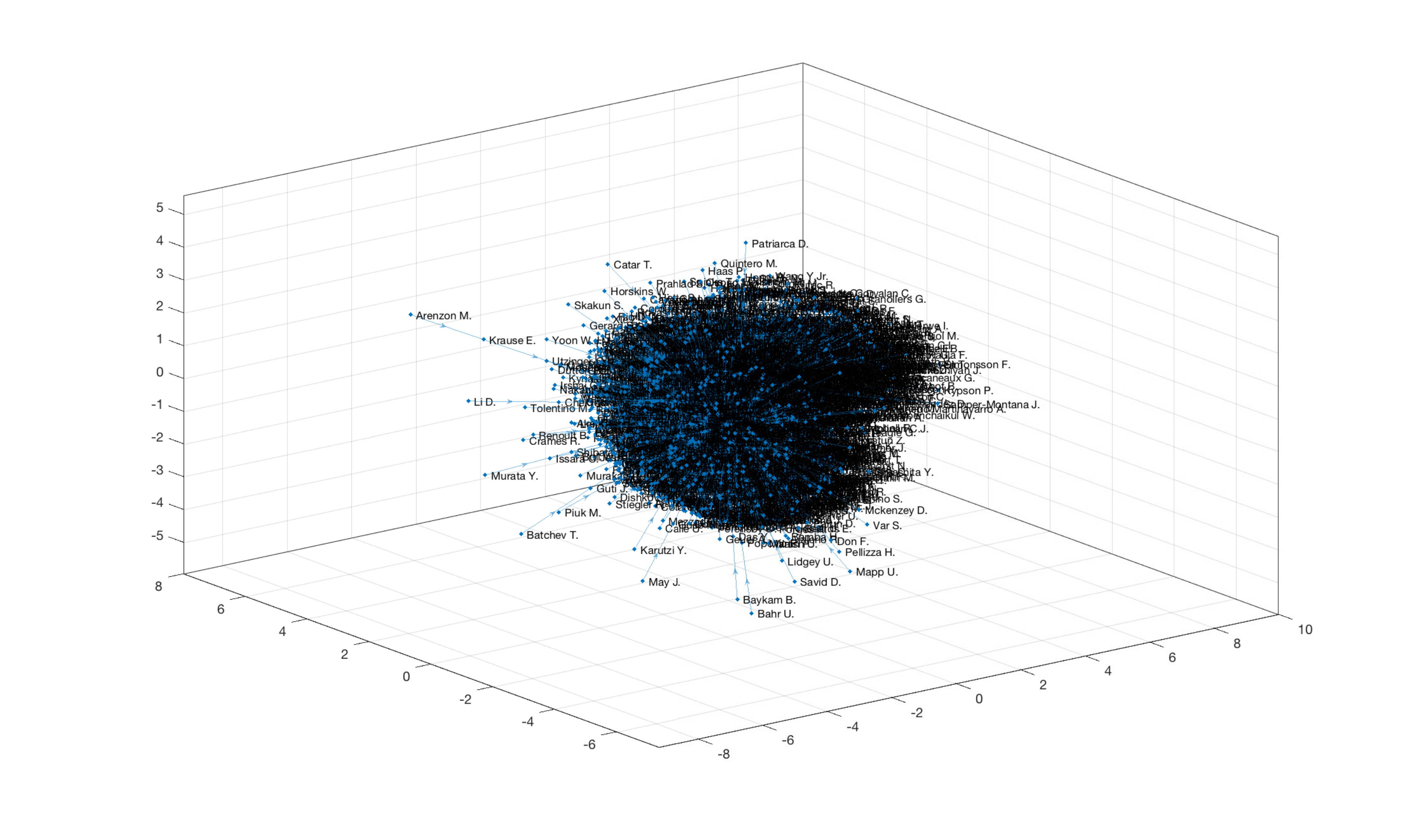}
\caption{Direct network visualization.}
\label{fig:network}
\end{figure}

In Figure \ref{fig:network} the visualization of the \textit{small} direct network is shown.

\begin{figure}[]
\hspace{-0cm}\includegraphics[width=\linewidth]{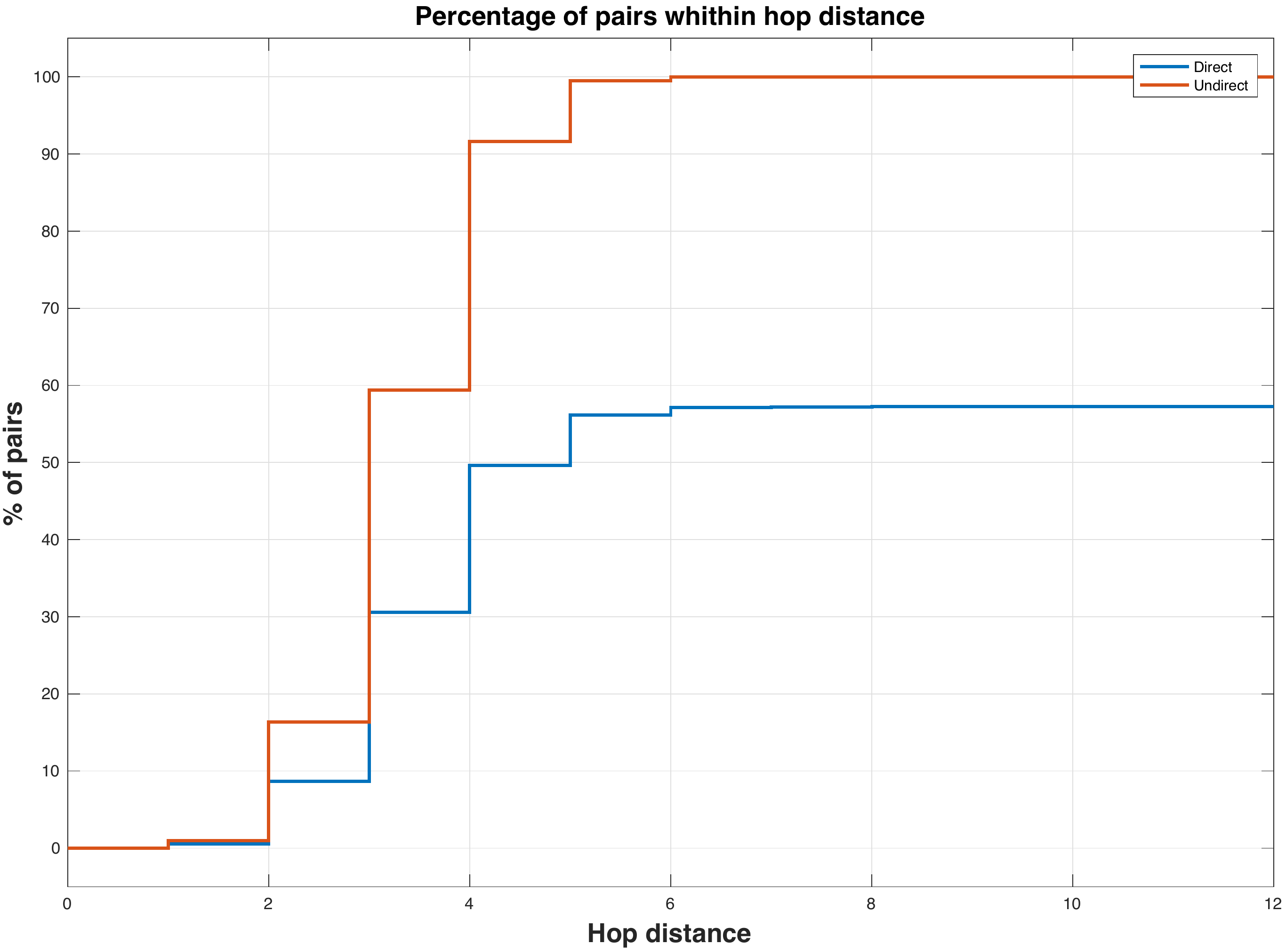}
\caption{Percentage of pairs within hop distance for unweighted directed and undirected \textit{small} networks.}
\label{fig:hops}
\end{figure}

By simply looking at the network topology and remembering how the network was constructed, we could already imagine that a giant component is present and that the small world property holds. By numerical evaluations, indeed, we found that in the direct representation there is one giant component of size $2428$ nodes and all the other components are unitary (in the undirect network there is just one component which contains all the nodes of the network). Defining the shortest path between any two nodes as the distance between those nodes we can derive the average distance of the direct graph defined as $\displaystyle \langle d\rangle=\frac{1}{N} \sum_{i=1}^N d_{i}$ with $d_i=\displaystyle \frac{1}{N} \sum_{j=1}^N \mathrm{min}(d_{ij})$ which lead to $\langle d\rangle\approx3.48$ hops (and for the \textit{small} undirect network is $\langle d\rangle\approx3.34$ hops, for the large undirect network is $\langle d\rangle\approx 3.64 $ hops). Moreover the diameter of the network, i.e. the maximum of all the shortest paths, is $diam=\max(\min(d_{ij}))=10$ hops (and $diam=8$ hops for the \textit{small} undirect network, $diam=9$ hops for the last network).\\
Hence, the networks exhibit actually a \textit{strong} small world property which leads to very short distances between any chosen pair of nodes. In order to better visualize it we could also look at the plots of the percentage of nodes within a considered directed hop distance as in Figure \ref{fig:hops} and we realize that the worst cases are achieved only by a small fraction of nodes, thus reducing the variance of this metric. Notice that the blue curve for the first direct graph does not reach $100\%$ because some of the nodes are disconnected from the giant component.

\subsection{Degree Distributions}

\begin{figure}[]
\hspace{-0cm}\includegraphics[width=\linewidth]{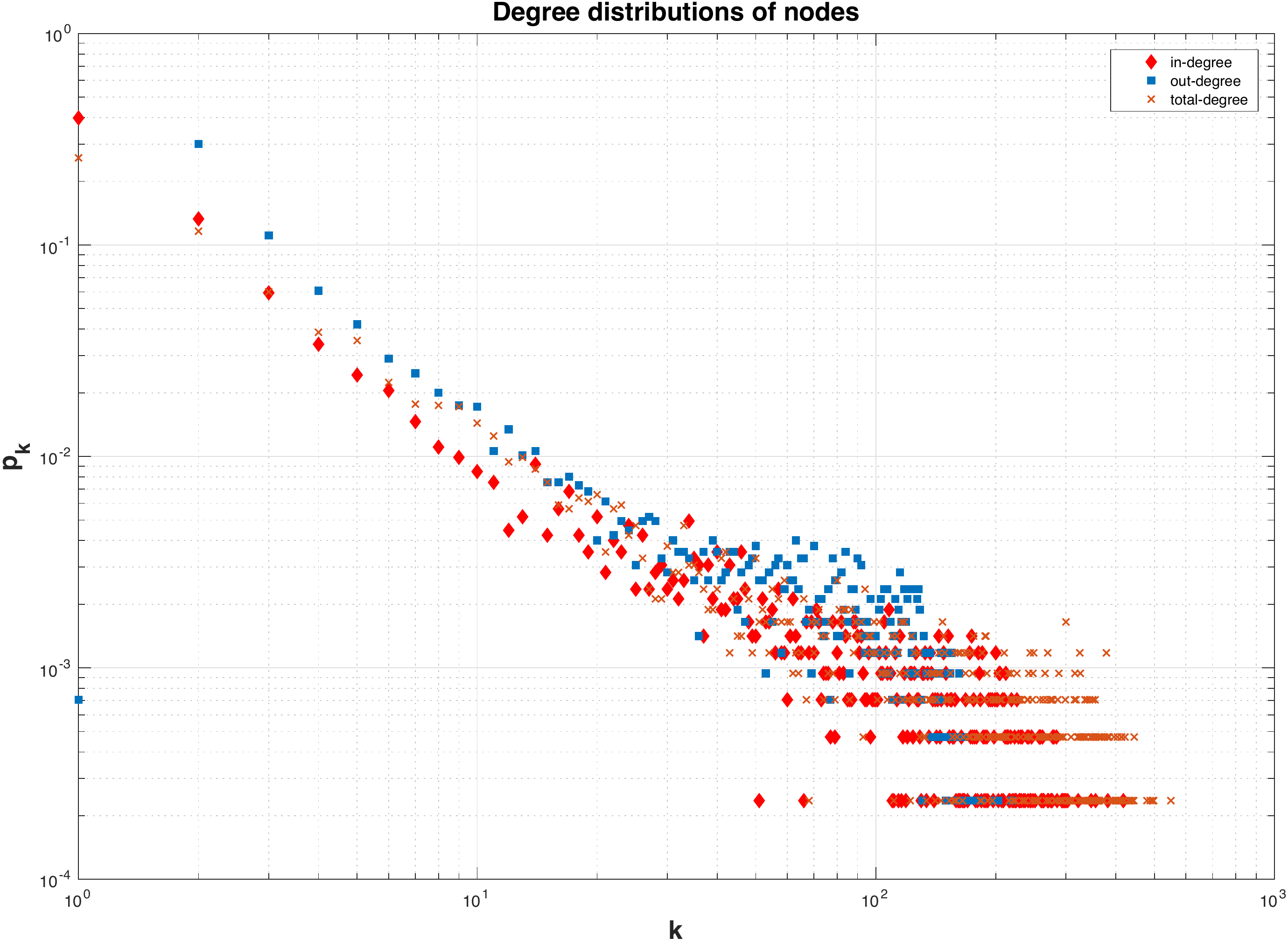}
\caption{Direct network degree distributions for in, out and total links in log-log scale.}
\label{fig:degree}
\end{figure}

Then it is interested the evaluation of the degree distributions of the nodes in the network. The distributions of in-degrees, of out-degrees and of total-degrees of the direct network are proposed in Figure \ref{fig:degree}. The values of $p_{in}(k)$, $p_{out}(k)$ and $p_{tot}(k)$ are the probabilities that a randomly picked node have $k$ incoming, outgoing or total links; i.e. the fraction of nodes that have in-degree, out-degree or total-degree equal to $k$. The distributions follow a similar behavior and we can notice that are heavy-tailed, thus there are some hubs in the network, i.e. outliers at high values, and we are going to explore them in the next section.

The average in-degree and out-degree are $\langle k \rangle \approx 23.89$ edges, and the total average degree is then the double $\langle k \rangle \approx 47.78$ edges. As already stated, the network is fully connected, which makes sense being $\langle k \rangle$ such a high value.

The second moment for in-degree is $\langle k_{in}^2 \rangle \approx 3.31 \cdot 10^3$, for out-degree is $\langle k_{out}^2 \rangle \approx 2.01 \cdot 10^3$ and for total-degree is $\langle k_{tot}^2 \rangle \approx 1.01 \cdot 10^4$.

\subsection{Hubs} \label{sec:res_hubs}

It has already been pointed out that the considered graphs present some hubs which are for interest in starting to determine the importance of a player in limitations to some specific metrics.\\
Indeed, the strongest players, identified as hubs, tend to play against a wide range of players: the \textit{weak} ones, generally at the first stages of the tournaments (the top-players, as tournament's seeds, are facilitated in the first rounds when they are called to face qualified players, which are generally weak), and the \textit{strong} ones at the last stages of the tournament, which are rarely reached by the weak players.\\
From the various typologies of network representation we can extract some useful information by simply looking at those hubs. 

\begin{figure}[]
\hspace{-0cm}\includegraphics[width=\linewidth]{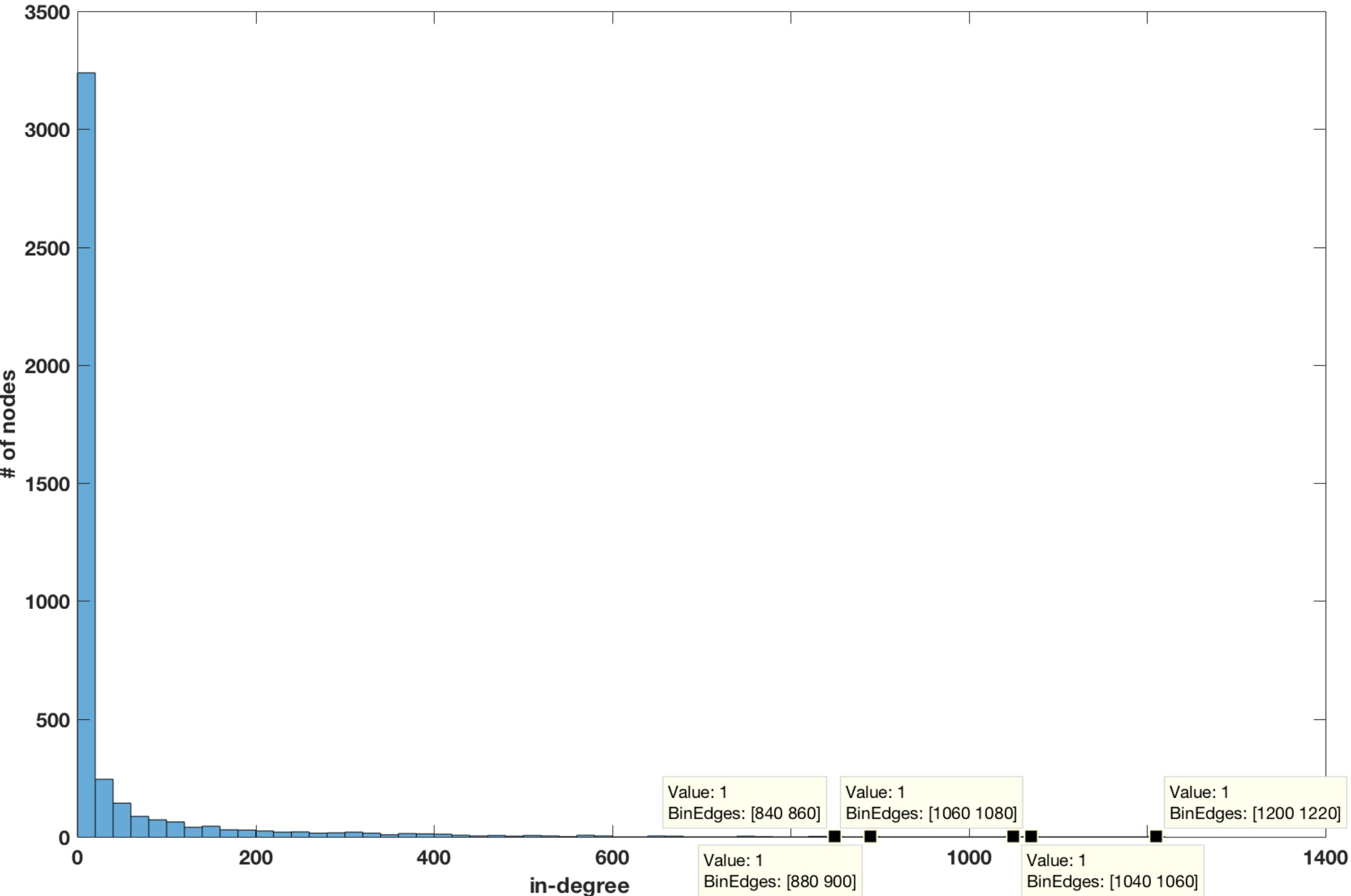}
\caption{Histogram of occurrences of in-degree nodes, in the x-label the index of the players.}
\label{fig:hist_datatips}
\end{figure}

First of all, in Figure \ref{fig:hist_datatips} we can see the histogram of occurrences of in-degree nodes in linear scale in order to be aware that the five highest-degree nodes are well spread apart from the majority of the other nodes which have much lower degrees.\\
Thus, if we look at the first five in-degree biggest hubs we can determine the players who won more matches and the respective number of winnings, those are: \textit{Jimmy Connors} (1219), \textit{Roger Federer} (1076), \textit{Ivan Lendl} (1047), \textit{Guillermo Vilas} (892) and \textit{John McEnroe} (840). If we compare with the ATP results archive we see slight variations, lower than 4\%, due to \textit{Davis Cup} matches.

Analogous procedures apply to the players who have lost the major number of matches: \textit{Fabrice Santoro} (436), \textit{Feliciano Lopez} (387), \textit{Mikhail Youzhny} (379), \textit{Guillermo Vilas} (373) and \textit{John McEnroe} (351). Notice that here the numbers are quite different from before, because only the players who also win a lot of matches are guaranteed to play in the major tournaments, otherwise after a while you loose the right to play in.

Surely the results seen up to now tend to promote already retired players and/or with a long career. On the other side, if we think of applying similar techniques for the network with ATP ranking weights, we see quite different outcomes promoting current-time players because nowadays there are great players, of course, but also more tournaments which give even more ATP points than in the previous generations of tennis players. Thus the five players who gained the most ATP points are: \textit{Roger Federer}, \textit{Rafa Nadal}, \textit{Novak Djokovic}, \textit{Jimmy Connors} and \textit{Ivan Lendl}. Unfortunately some inaccuracies are present in the translation of current ATP points for old matches.

Looking at the degree distributions in Figure \ref{fig:degree} and at the results we are suggested to consider an underlying assumption: the more connected athletes are and the most likely is to be best players. Most of the players have a small number of matches and then quit playing the major tournaments, on the contrary, there is a small group of top-players who perform many matches against weaker players and among themselves.
This phenomenon is an observation of the \textit{rich get richer} effect driven by the attractiveness of the high connected nodes as opponent for new-comers; an interpretation of the \textit{richness} that the players achieve could be their gain of some sort of "experience" during the matches of their carreer, as already pointed out in \cite{situngkir}.

\subsection{Considerations on Network Nature} \label{sec:res_nature}

\begin{figure}[]
\hspace{-0cm}\includegraphics[width=\linewidth]{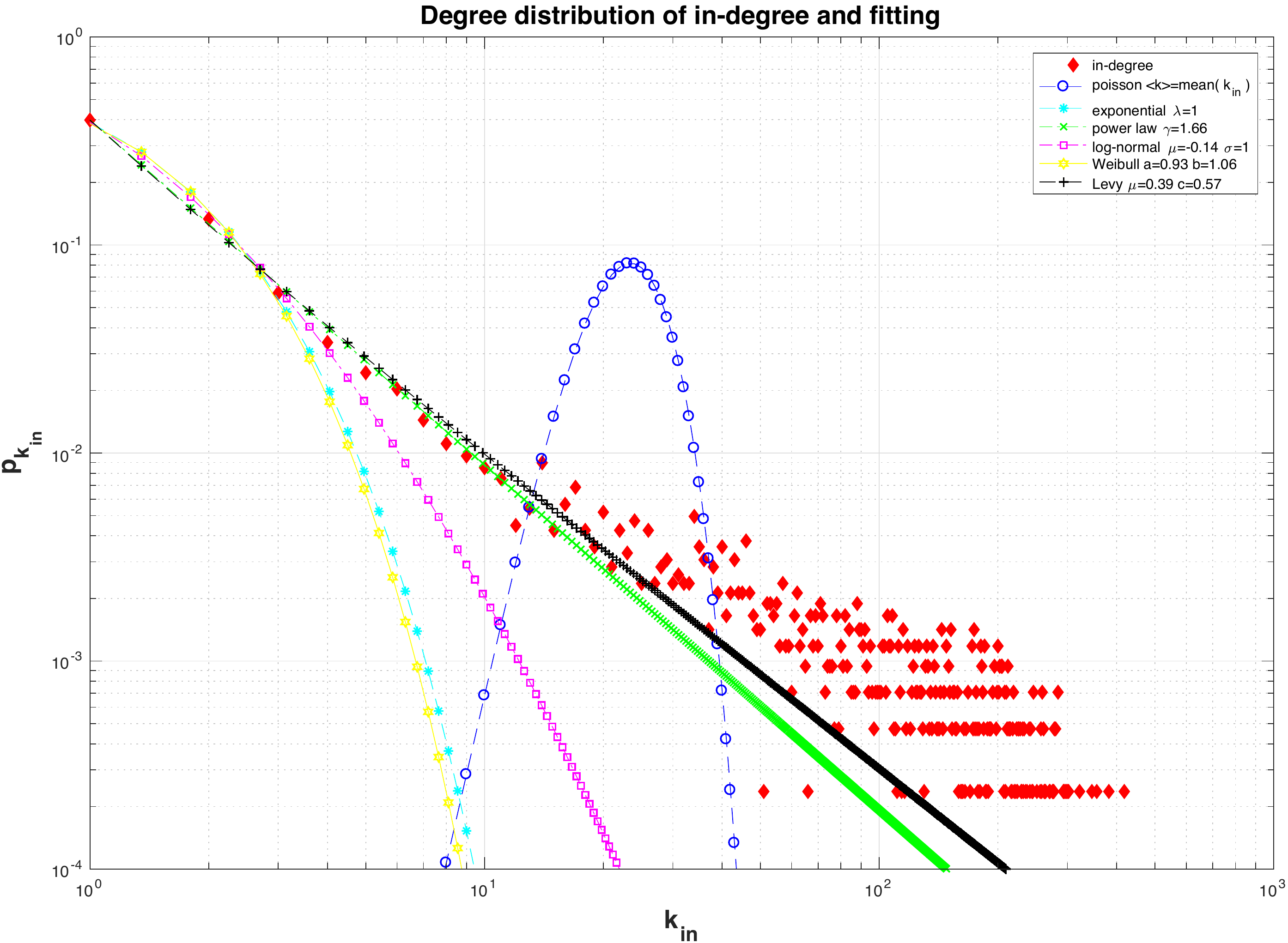}
\caption{Fitting trials of raw data of in-degrees, log-log scale.}
\label{fig:fit}
\end{figure}

The most critical point in the analysis already present in literature on this topic is the scale-free nature of such networks. A first-step analysis about the network nature can be done by plotting again the degree distribution (e.g. the in-degree) and trying to fit it with some typical network distributions. The results are shown in Figure \ref{fig:fit} where we can appreciate the differences among them. The respective parametric formulations of distributions together with fitting parameters and the coefficient of determination $R^2$ are reported in Table \ref{tab:fit}. As already noticeable by the presence of hubs, the network cannot clearly follow a random model (the poisson one) and some heavy-tailed distributions need to be considered. 
The power-law and the Lévy distributions are the two models performing better on the raw data, thus the considered network exhibits many properties typical of scale-free networks.\\
Assuming the network as power-law, we can measure the scale-free parameter $\gamma$: it is found to be $\gamma_{in}\approx 1.66$ for in-degrees, $\gamma_{out} \approx 2.12$ for out-degrees and $\gamma_{\mathrm{undirect}} \approx 1.38$ for the undirect network, values which are consistent with the ones found in literature in \cite{radicchi} and \cite{situngkir}. 

\setlength{\tabcolsep}{0.9em} % for the horizontal padding
{\renewcommand{\arraystretch}{1.2}
\begin{table*}[]
\centering
\begin{tabular}{l|l|l|l|l|}
\cline{2-5}
                                  & pmf or PDF                                                                                    & \multicolumn{2}{l|}{Numerical parameters} & $R^2$ \\ \hline
\multicolumn{1}{|l|}{Poisson}     & $\displaystyle p_k = e^{-\lambda} \frac{\lambda^k}{k!}$                    &   $\lambda=\langle k \rangle$                  &                     &       \\ \hline
\multicolumn{1}{|l|}{Exponential} & $\displaystyle p_k=\lambda e^{-\lambda k}$   & $\lambda=1$                 &                     &    $0.9879$   \\ \hline
\multicolumn{1}{|l|}{Power-law}   & $\displaystyle p_k \propto k^{-\gamma}$    &        $\gamma=1.66$                             &                     & $0.998$      \\ \hline
\multicolumn{1}{|l|}{Log-normal}  & $\displaystyle p_k= \frac{1}{x \sigma \sqrt{2 \pi}}  e^{-\frac{(\ln(x)-\mu)^2}{2 \sigma^2}} $   &   $\mu=-0.14$               &   $\sigma=1$             &  $0.9938$     \\ \hline
\multicolumn{1}{|l|}{Weibull}     & $\displaystyle p_k = a b x^{(b-1)} e^{-a*x^b}$         &      $a=0.93$            &         $b=1.06$        &    $0.9866$   \\ \hline
\multicolumn{1}{|l|}{Lévy}        & $\displaystyle p_k = \sqrt{\frac{c}{2 \pi}}  \frac{e^{-\frac{c}{2 (x-\mu)}}}{(x-\mu)^{3/2}} $ &  $\mu=0.39$       &  $c=0.57$           &   $0.9979$    \\ \hline
\end{tabular}
\vspace{0.2cm}\caption{Fitting distribution applied to in-degree raw data.}
\label{tab:fit}
\end{table*}
}

The aforementioned results need to be taken with some cautions; the network characteristics are quite similar to scale-free networks (they are even more similar when we restrict our interest to top-players and/or to top-tournaments only) and scaling behavior is also suggested by the structural \textit{preferential attachment} of new players who generally tend to connect to an existing player with a probability proportional to the degree of such node. 
However if we try to reduce the noise around the outliers, e.g. by considering the cumulative degree distribution or a log-binning of the data, we can see that the network is not a pure scale-free model and it would make sense to limit the intervals introducing cutoff values $k_{min}$ and $k_{max}$, as carefully proved in \cite{delgenio} in a general setting. Moreover, as suggested in \cite{fitting} we should not rely on the $R^2$ parameter, since it is proved to achieve very high values also for non scale-free networks. Summing up: the fit on raw data based on $R^2$ is source of many errors in current literature about network analysis. 

\begin{figure}[]
\hspace{-0cm}\includegraphics[width=\linewidth]{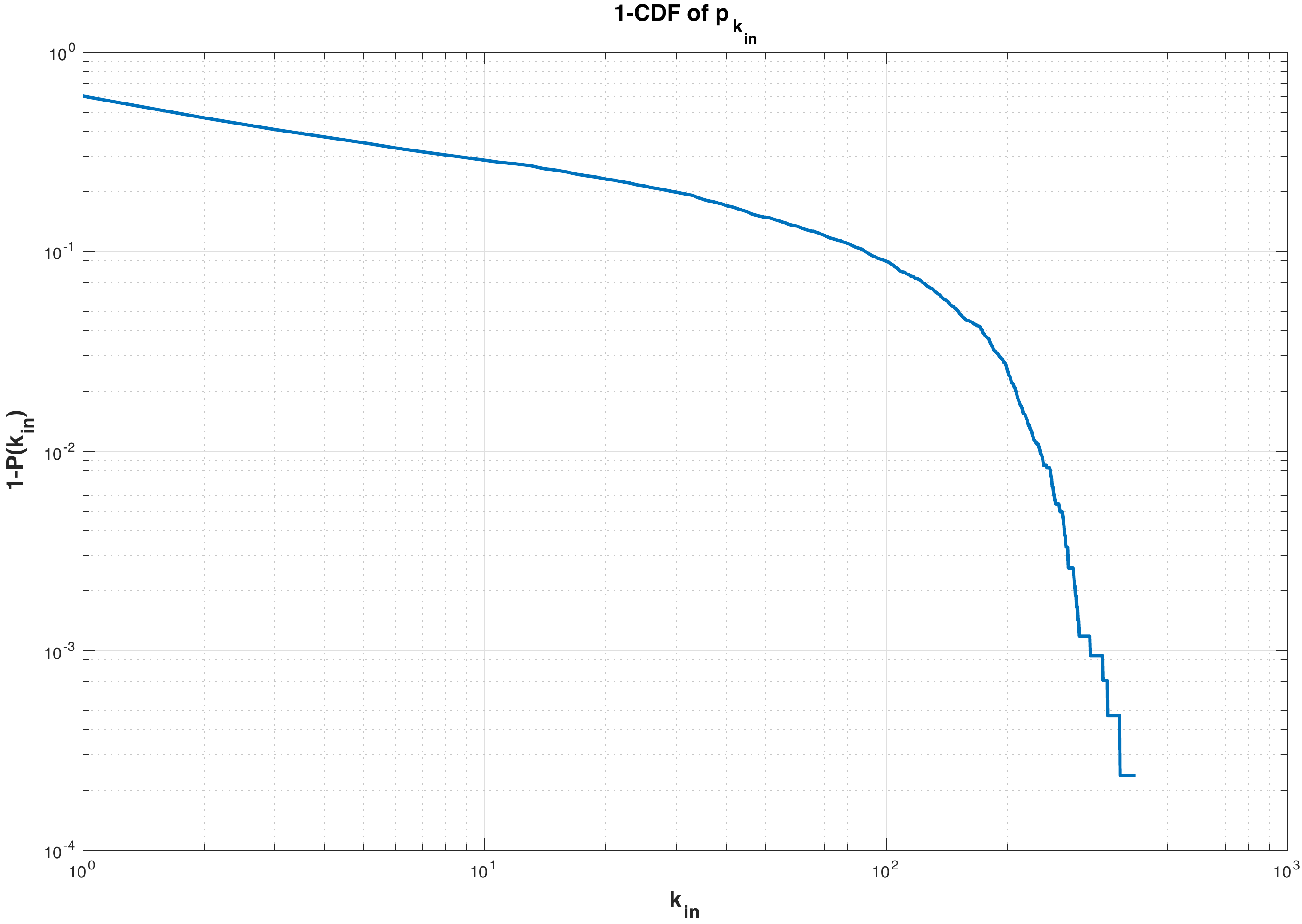}
\caption{Complementary Cumulative Distribution Function (CCDF) in log-log plot.}
\label{fig:cumulative}
\end{figure}

Hence, the Complementary Cumulative Distribution Function (CCDF) should be considered and it is plotted in Figure \ref{fig:cumulative}; here the issue of the plateau corresponding to values occurring once has been solved and we can confirm the previous considerations since power-law networks would describe a straight line in a log-log plot.\\ More specifically, replicating the fitting on the cumulative distribution, the network can be categorized as a power-law with an exponential cut-off, i.e. $\displaystyle p_k \propto x^{-\alpha} e^{-\frac{k}{\beta}}$: indeed we can observe in Figure \ref{fig:cumulative} that the curve starts out as a power law and ends up as an exponential. \\
This result confirms that $R^2$ on raw data cannot be trusted and that, maybe, often is just worth noticing that the distribution has a heavy tail instead of asserting immediately its scale-free nature, which is a very widespread practice (many interesting considerations about networks and fitting techniques can be read in \cite{fitting}).\\
The fact that the complete network is not scale-free is a quite surprising result, although very clear from the CCDF plot, because all other related studies on tennis network are affirming the scale-free nature and are deriving from there the interesting properties of the network, which is not precisely correct. \\
Some power-law networks with coefficient lower than 2 or not scale-free networks at all have gained a lot of attention in recent literature, for example in \cite{facebook} and \cite{seyed}, because those models have been discovered in many real scenarios where the number of new links generally grows faster than the number of new nodes, which is precisely our situation. The number of annual tennis matches is nowadays very big and therefore the probability of a new connection is much more frequent than the new players who join the most important ATP tour. Thus it may seem that the network should evolve toward a non-sparse adjacency matrix, which is still not the case because of structural constraints: is very unlikely that some pairs of players will face against each other (because of players' retirement from the tour or players which are strongly far apart in ranking) but on the other hand the hubs role of long-career players will be even reinforced.

\subsection{Clustering Coefficient}

\begin{figure}[]
\hspace{-0cm}\includegraphics[width=\linewidth]{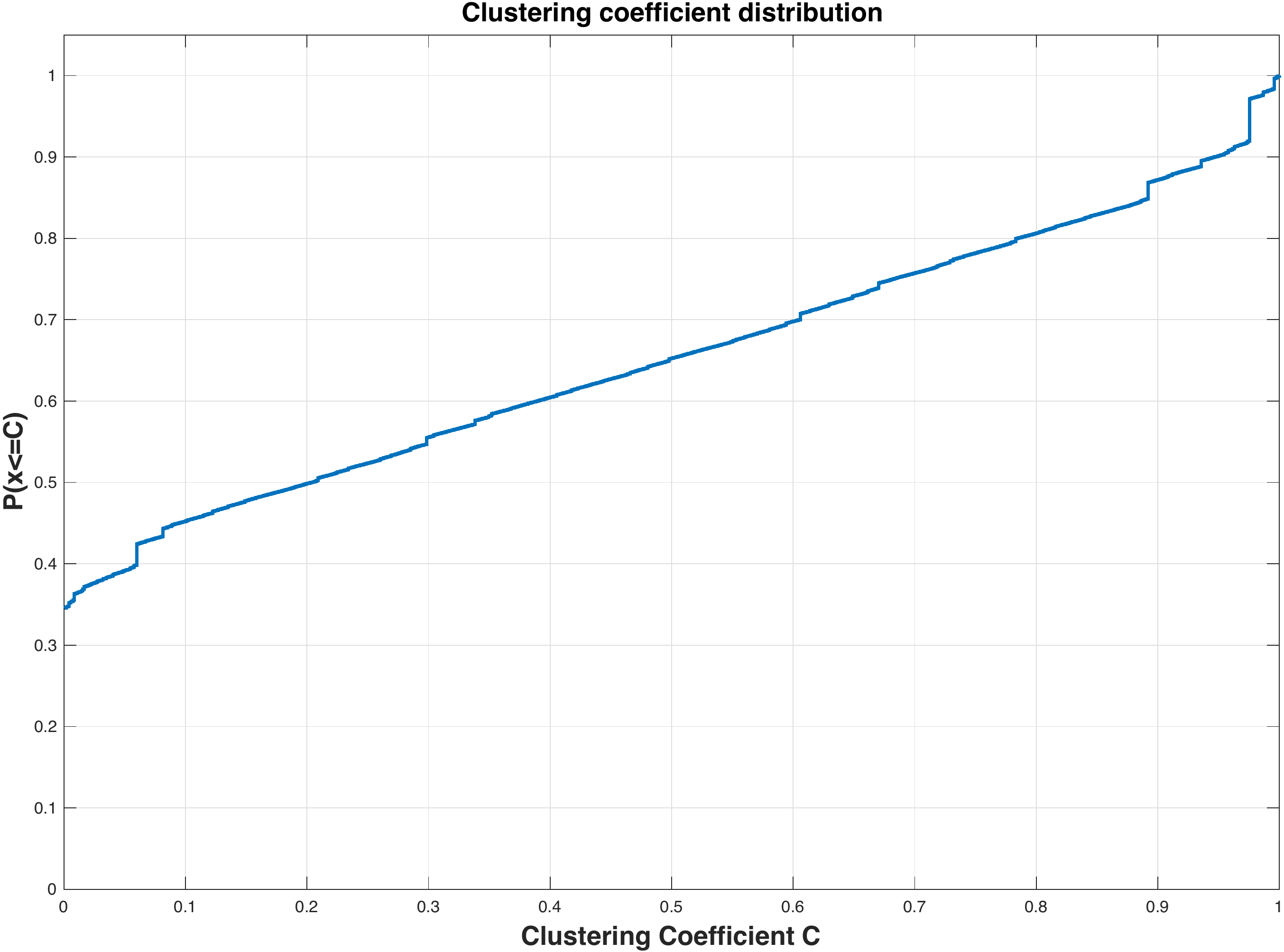}
\caption{Clustering coefficient distribution of the undirect representation.}
\label{fig:clustering}
\end{figure}

Another interesting property of small world network is the clustering coefficient $C$. For a node $j$ its clustering coefficient, $C_j$, is a number belonging to $[0, 1]$ denoting how many links there are between its neighboring nodes normalized to the maximum possible number of links among them; more formally could be defined as $C_j=E_j/E_{j, \mathrm{max}}$ where $E_j$ is the number of edges between nodes in the neighborhood of $j$ ($\mathcal{N}_j$) and $E_{j,\mathrm{max}}$ is its maximum value. \\
For example, in undirect networks it holds $\displaystyle 0\le E_j \le E_{j,\mathrm{max}}=|\mathcal{N}_j|(|\mathcal{N}_j|-1)/2$. Finally, the general clustering coefficient is expressed as the average over all the $N$ players: $\displaystyle C=\frac{1}{N}\sum_{j=1}^N C_j$.\\
In the undirect representation of the network we have found $C=0.07$, which is coherent as order of magnitude with the values found in \cite{situngkir}; as a term of comparison, in Figure \ref{fig:clustering} is reported the clustering coefficient distribution, i.e. the fraction of nodes having clustering coefficient lower than $c$.

\subsection{Degree Correlation}

Another interesting metric on the network point of view is the correlation between the degrees of nodes.
Degree correlation can be expressed in many ways, usually the Pearson assortativity coefficient is used as principal investigation method and is defined as: $$r=\frac{\sum_{h=\min(k)}^{\max(k)}   \sum_{f=\min(k)}^{\max(k)}  h f (e_{hf}-q_h q_f)  }{\sigma^2}$$ where $q_h$ is the probability of finding a degree-$h$ node at the end of a randomly picked link, $e_{hf}$ is the probability of finding a link between two nodes of degree $h$ and $f$, $\sigma ^2$ is the variance of the degrees and can be proved to be the maximum of the numerator, thus $r\in[-1, 1]$. Computing this parameter for the \textit{small} undirect and unweighted network we found $r=-0.0076$ which means a slightly assortative network and actually almost neutral, probably because hubs and small degree nodes are likely to play against in the first stages of the tournaments but also hubs tend to play against themselves in the final rounds, thus no strong pattern exists. For the large undirect network $r= 0.272 $ is found.

%In Table \ref{tab:stats} some statistics are summarized for the two undirected and unweighted networks.

%\begin{table}[]
%\centering
%\caption{Topological measures of two networks presented.}
%\label{tab:stats}
%\begin{tabular}{lcc}
%                 & Network \#3 & Network \#4 \\
%$N$               & 4245        & 22405       \\
%$L$               & 85084       & 499057      \\
%$<k>$             & 40.087      & 44.549      \\
%density           & 0.009       & 0.002       \\
%$C$               & 0.07       & 0.12       \\
%charpath          & 3.340       & 3.639       \\
%eff glob          & 0.326       & 0.293       \\
%eff loc           & 0.434       & 0.286       \\
%assortativity     & -0.008      & 0.272       \\
%powerlaw $\gamma$ & 1.380       & 1.380      
%\end{tabular}
%\end{table}

\subsection{Robustness to Failures}

\begin{figure}[h]
\hspace{-0cm}\includegraphics[width=\linewidth]{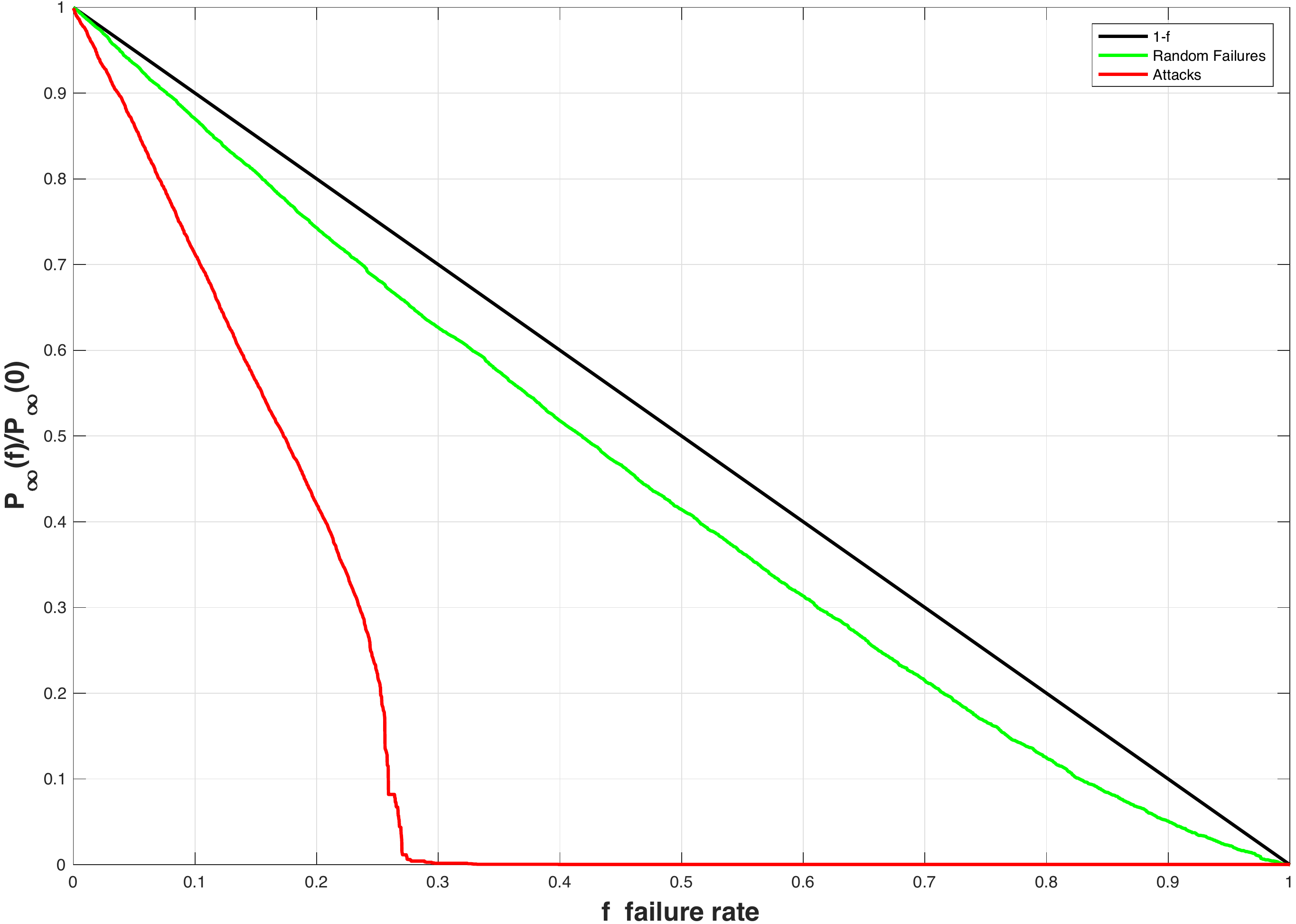}
\caption{Robustness of the undirect network after $f\%$ of nodes removals.}
\label{fig:failures}
\end{figure}

For a more complete characterization of the network structure it could be interested to analyze its robustness to failures, i.e. the nodes removal from the network. In the chosen network a player could be disqualified for doping or other reasons, or we could need to consider just a subset of players or matches (restricting by nationality, left or right handedness, height threshold, tournament level and so on). We want to determine the robustness of the network in terms of percentage of nodes connected to the giant component when $f\%$ of its nodes has been removed. In the following we are just considering \textit{random} removals and \textit{attack-based} removals since all the other are mainly application-driven and can be done with a small effort manipulating the dataset as desired. In the first scenario considered the nodes are removed entirely at random while in the second scenario the highest hub in the network is removed at each step. We introduce the probability that a random node belongs to the giant component after that $f\%$ of nodes have been removed as $P_{\infty}(f)$ and we can look at the relative size of the giant component: $P_{\infty}(f)/P_{\infty}(0)$, where $P_{\infty}(0)$ represents the best case of no removals thus the ratio belongs to $[0, 1]$. The plot of such ratio for the undirect graph is shown in Figure \ref{fig:failures} and we recognize in our network the high robustness typical to well-connected and scale-free graphs. The black line corresponds to $1-f$ and it is an upper limit since for sure we have removed $f\%$ of nodes from the network (and then also from the giant component).

\subsection{Ranking Methods and Predictive Power} \label{sec:res_ranking}

The main interest in analysis of tennis networks is directed towards the implementation of new ranking techniques instead of the ATP ranking system, together with their relative predictive power when a new match is played. There are no contributions in literature, instead, for what concerns link prediction (e. g. who are the most probable players to play against given that they did not played against before?) and community detection. \\ %(possibly overlapping)
Radicchi in \cite{radicchi} is the first who applied the PageRank (PR) algorithm to tennis network thus identifying \textit{Jimmy Connors} as the most valuable player in the tennis history. Dingle et al. in \cite{dingle} applied the previous work to ATP and WTA (\textit{Women's Tennis Association}) matches and they also provided a simple comparison on the basis of predictive power. In \cite{subgraph} the authors proposed yet another ranking method applying PageRank to the subgraph of the Top-100 players. In \cite{spanias} many ranking methods, both through network and Markov chains analysis, have been proposed and verified by means of prediction power.

Also few other non-network-related approaches have been proposed so far: for example in \cite{developing} statistical models have been tested to improve the current ranking system, or in \cite{ML} the authors applied a novel method exploiting neural networks based on 22 features and achieving a 75\% benefit in prediction through those techniques.

\subsubsection{Preliminaries on Ranking Algorithms}
The analysis shown in this section assumes the direct representation of the network where the loser player has an edge to the winner player and the weight corresponds to the number of matches won by the winner. \\
The link analysis methods investigated in this report are: \textit{Hubs and Authorities} (HITS algorithm, \textit{Hyperlink-Induced Topic Search} discussed in \cite{hits}), simple PageRank and PageRank with teleportation (see \cite{PR} and \cite{PR_teleportation} as references). 

The idea on which HITS algorithm is based regards the definitions of hubs and authorities. Authorities are nodes with a high number of edges pointing to them, hubs are nodes which link to many authorities; in our scenario, intuitively, we expect that authorities are often associated with the most successful players (because they won against a wide gamma of players), while the hubs with \textit{mediocre} players with a long career. More formally we can compute the authority-scores $\mathbf{a}$ and the hub-scores $\mathbf{h}$ respectively as: $$\displaystyle \mathbf{a}=\frac{\mathbf{A h}}{||\mathbf{A h}||} \quad \quad \quad \quad \mathbf{h}=\frac{\mathbf{A^T a}}{||\mathbf{A^T a}||}$$ where we assumed the adjacency matrix to be the transposed version of the one presented above (i.e. here an entry $a_{ij}=1$ means that there is a link from node $j$ to node $i$). Notice that $a_i\geq0$ and $h_i\geq0$ $\forall i$. The problem can be solved through power iteration with convergence parameter $\epsilon$.

The rationale behind PageRank is that of a random walk along the graph and the \textit{prestige} score $\mathbf{p}$ for each player is determined as the probability of being at that node in stationarity conditions. The t-th update of $\mathbf{p}$ goes as: $$\mathbf{p}_t=\mathbf{M p}_{t-1}$$ where $\mathbf{M}$ is the column stochastic adjacency matrix.

The simple PageRank algorithm is affected by some undesirable problems. For example it would end up in dead ends, although it is not the case because who win one match will surely lose one other (unless the player plays only tournaments winning all of them, which never happens); and there also might be periodic behavior looping in cycles, which is somehow reasonable to expect since we are considering very different tennis epochs. Thus we can add to the model the possibility of not to follow the behavior but to jump to a random node in the network with a probability $\alpha\in (0,1)$. Hence the t-th update step of $\mathbf{p}$ becomes: $$\mathbf{p}_t=\mathbf{\breve{M} p}_{t-1}=(1-\alpha)\mathbf{q_1 1^T p}_{t-1}+\alpha \mathbf{M p}_{t-1}$$ where $\mathbf{q_1}$ is the stochastic teleportation vector and we assumed it to be $\mathbf{q_1}=\frac{1}{N}\mathbf{1}$ (equal probabilities), with $\mathbf{1}$ column vector of N ones; $\alpha$ is a damping factor typically set to $0.85$ (this is due to historical reasons as proposed in the original paper \cite{PR_teleportation} and for the sake of comparisons with other works). This considerations let to write a much simpler iteration procedure than the one proposed in \cite{radicchi} and \cite{dingle}, although they are equivalent. The simplifications are made possible thanks to the observation regarding the absence of sinks-like nodes and to a compact vectorial expression.

\subsubsection{Discussion of results}
\begin{figure}[]
\centering
\hspace{0cm}\includegraphics[width=\linewidth]{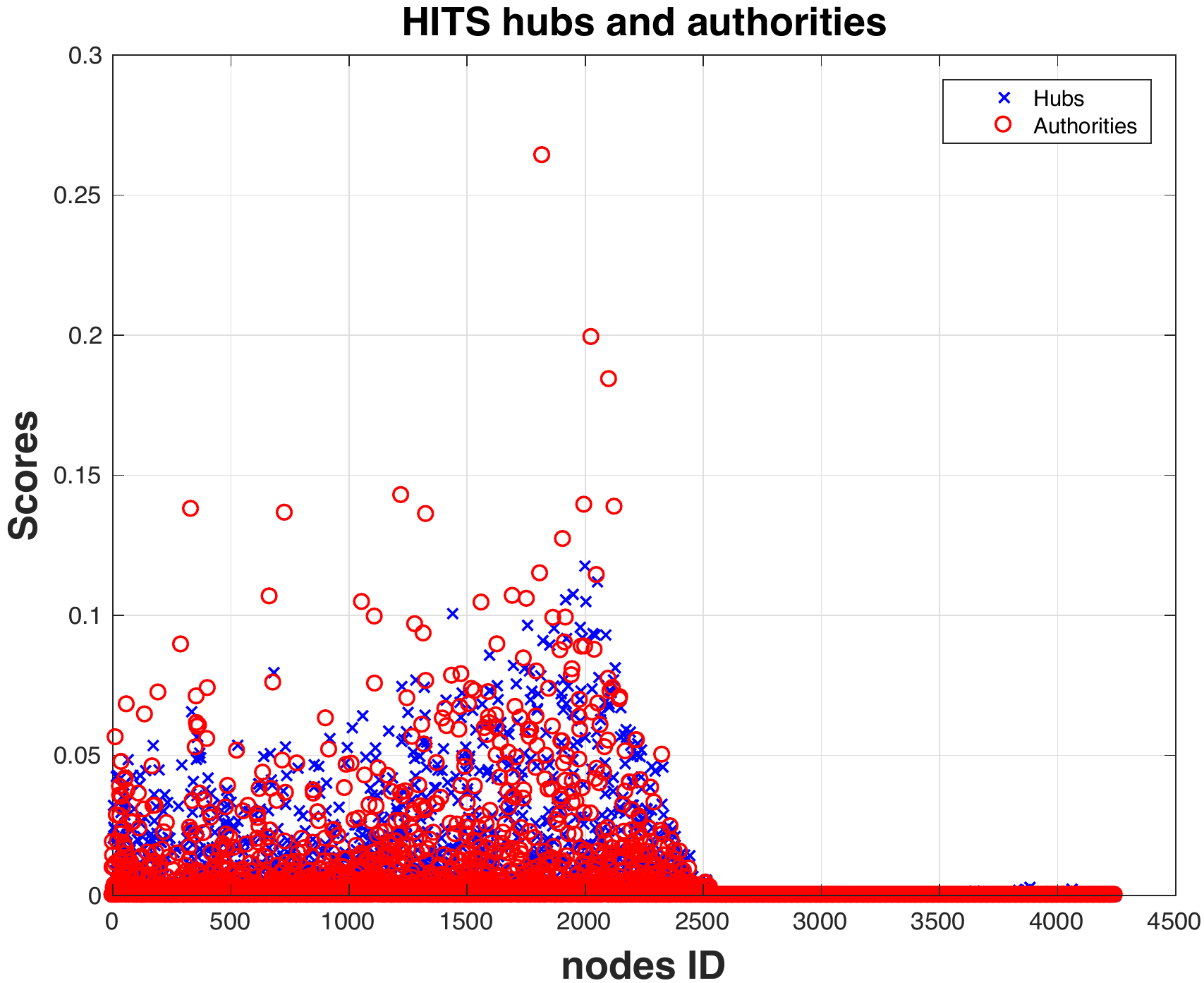}
\vspace{-0.7cm}\caption{Hubs and authorities scores of HITS algorithm.}
\label{fig:hits}
\end{figure}

Hubs and authorities scores are reported in Figure \ref{fig:hits}, where can be seen that nodes ID corresponding to players who only have lost matches (the last ones) have zero authority score, because there are no links pointing to them, but possibly non-zero hub score.\\
The names of the Top-20 hubs and authorities are reported in the second and third columns of Table \ref{tab:rankings} for $\epsilon=10^{-8}$; few changes happen varying this parameter and most of them not in the very first positions. Though there are no reference literature of HITS applied to tennis, nevertheless from the table we can confirm our previous intuition and also realize that those concepts are somehow similar to what already discussed talking about in-degree and out-degree hubs. Indeed, we can recognize that in-degree hubs and out-degree hubs are placed in the first positions respectively of authorities and of hubs, although not in the precise order. Compare these results with section \ref{sec:res_hubs}

\begin{table*}[]
\hspace{-0.5cm}
\renewcommand{\arraystretch}{1.33}
\renewcommand{\tabcolsep}{1.23mm}
\begin{tabular}{|c|c|c|c|c|c|c|c|}
\hline
\textbf{Rank} & \textbf{Authorities}                          & \textbf{Hubs}                            & \makecell{\textbf{Simple}\\ \textbf{PR}}                  & \makecell{\textbf{PR with}\\ \textbf{Teleport.}}  & \makecell{\textbf{Authorities}\\\textbf{2017}}  & \makecell{\textbf{PR with}\\ \textbf{Teleport. 2017}} & \makecell{\textbf{Official ATP}\\ \textbf{2017}}  \\ \hline
1    & \textit{\textbf{Roger Federer}}      & \textit{David Ferrer}           & \textit{\textbf{Roger Federer}}  & \textit{\textbf{Jimmy Connors}} & \textit{\textbf{\underline{Rafael Nadal}}} &\textit{\textbf{Roger Federer}}&\textit{\textbf{Rafael Nadal}}\\ \hline
2    & \textit{\textbf{Rafael Nadal}}       & \textit{Tomas Berdych}          & \textit{\textbf{Rafael Nadal}}   & \textit{\textbf{Ivan Lendl}}   & \textit{\textbf{\underline{Roger Federer}}} &\textit{\textbf{Rafael Nadal}}&\textit{\textbf{Roger Federer}}\\ \hline
3    & \textit{\textbf{Novak Djokovic}}     & \textit{Feliciano Lopez}        & \textit{\textbf{Novak Djokovic}} & \textit{\textbf{Roger Federer}} & \textit{Alexander Zverev}&\textit{Alexander Zverev}&\textit{Grigor Dimitrov}\\ \hline
4    & \textit{\textbf{Andre Agassi}}       & \textit{Mikhail Youzhny}        & \textit{\textbf{Ivan Lendl}}     & \textit{\textbf{John McEnroe}}  & \textit{Grigor Dimitrov}&\textit{David Goffin}&\textit{Alexander Zverev}\\ \hline
5    & \textit{David Ferrer}                & \textit{Fernando Verdasco}      & \textit{\textbf{Andre Agassi}}   & \textit{\textbf{Rafael Nadal}}  & \textit{David Goffin}& \textit{Grigor Dimitrov}&\textit{Dominic Thiem}\\ \hline
6    & \textit{\textbf{Andy Murray}}        & \textit{Fabrice Santoro}        & \textit{\textbf{Pete Sampras}}   & \textit{\textbf{Novak Djokovic}} & \textit{Dominic Thiem}&\textit{J. M. Del Potro}&\textit{Marin Cilic}\\ \hline
7    & \textit{\textbf{Jimmy Connors}}      & \textit{Tommy Haas}             & \textit{\textbf{Andy Murray}}    & \textit{Guillermo Vilas}     &   \textit{Marin Cilic} &\textit{Dominic Thiem}&\textit{David Goffin}\\ \hline
8    & \textit{\textbf{Ivan Lendl}}         & \textit{Jarkko Nieminen}        & \textit{\textbf{Jimmy Connors}}  & \textit{\textbf{Ilie Nastase}}  & \textit{\underline{Jack Sock}}&\textit{\underline{Jack Sock}}&\textit{Jack Sock}\\ \hline
9    & \textit{\textbf{Pete Sampras}}       & \textit{Tommy Robredo}          & \textit{David Ferrer}            & \textit{\textbf{Andre Agassi}} &  \textit{Roberto B. Agut}&\textit{Nick Kyrgios}&\textit{Stan Wawrinka}\\ \hline
10   & \textit{\textbf{Andy Roddick}}       & \textit{Philipp Kohlschreiber}  & \textit{\textbf{Stefan Edberg}}  & \textit{\textbf{Bjorn Borg}}   & \textit{J. M. Del Potro} &\textit{Marin Cilic}&\textit{Pablo C. Busta}\\ \hline
11   & \textit{\textbf{Lleyton Hewitt}}     & \textit{Andreas Seppi}          & \textit{\textbf{Boris Becker}}   & \textit{\textbf{Stefan Edberg}} & \textit{Pablo C. Busta} &\textit{Sam Querrey}&\textit{J. M. Del Potro}\\ \hline
12   & \textit{Tomas Berdych}               & \textit{Stanislas Wawrinka}     & \textit{\textbf{Andy Roddick}}   & \textit{\textbf{Pete Sampras}} & \textit{Diego Schwartzman}  &\textit{Roberto B. Agut}&\textit{\textbf{Novak Djokovic}}\\ \hline
13   & \textit{\textbf{Carlos Moya}}        & \textit{Richard Gasquet}        & \textit{\textbf{John McEnroe}}   & \textit{\textbf{Arthur Ashe}}    & \textit{Lucas Pouille}&\textit{Jo-Wilfried Tsonga}&\textit{Sam Querrey}\\ \hline
14   & \textit{\textbf{John McEnroe}}       & \textit{Nikolaj Davydenko}      & \textit{\textbf{Lleyton Hewitt}} & \textit{\textbf{Boris Becker}}  & \textit{Tomas Berdych}&\textit{Giles Muller}&\textit{Kevin Anderson}\\ \hline
15   & \textit{Tommy Haas}                  & \textit{\textbf{Roger Federer}} & \textit{Tomas Berdych}           & \textit{Stan Smith}            & \textit{\underline{Jo-Wilfried Tsonga}}& \textit{\textbf{Novak Djokovic}}&\textit{Jo-Wilfried Tsonga}\\ \hline
16   & \textit{\textbf{Stefan Edberg}}      & \textit{Radek Stepanek}         & \textit{Michael Chang}           & \textit{Brian Gottfried}   & \textit{\textbf{Novak Djokovic}} &  \textit{Tomas Berdych} &\textit{\textbf{Andy Murray}}\\ \hline
17   & \textit{\textbf{Yevgeny Kafelnikov}} & \textit{Jonas Bjorkman}         & \textit{\textbf{Yevgeny Kafelnikov}}      & \textit{Manuel Orantes}      & \textit{Milos Raonic}  & \textit{Milos Raonic}&\textit{John Isner} \\ \hline
18   & \textit{\textbf{Boris Becker}}       & \textit{\textbf{Carlos Moya}}   & \textit{Goran Ivanisevic}        & \textit{\textbf{Andy Murray}} &  \textit{Philipp Kohlschreiber} &\textit{Kevin Anderson}&\textit{Lucas Pouille}\\ \hline
19   & \textit{Nikolaj Davydenko}           & \textit{\textbf{Andy Murray}}   & \textit{\textbf{Carlos Moya}}    & \textit{David Ferrer}      &  \textit{Kevin Anderson}   & \textit{Damir Dzumhur}&\textit{Tomas Berdych}\\ \hline
20   & \textit{Tommy Robredo}               & \textit{Ivan Ljubicic}          & \textit{Tommy Haas}              & \textit{Roscoe Tanner}       &  \textit{John Isner}& \textit{Alberto R. Vinolas} &\textit{Roberto B. Agut}\\ \hline
\end{tabular}
\bigskip
\vspace{-0.2cm}\caption{Ranking methods outcomes; the bold names are players who have been at the first ATP position during their career. Players like \textit{Manuel Orantes, Guillermo Vilas} and \textit{David Ferrer} are often referred to as eternal second best and in the collective imagination they deserved to be number one of the ranking. Underlined names in the last columns are the ones ranked in the same position as in Official ATP ranking.}
\label{tab:rankings}
\end{table*}

In terms of complexity we expect at most $t_{\max}$ iterations for the HITS algorithm to converge, where:
$$t_{\max}=\ceil[\Bigg]{   - \frac{\ln(\epsilon)-\ln(\sqrt{N})}{2 \ln(d_1/d_2)}       }$$ with $d_1$ and $d_2$ being the eigenvalues associated with the two highest eigenvectors of $\mathbf{M}=\mathbf{AA^T}$. Setting $\epsilon=10^{-8}$ it results in $t_{\max}=180$ iterations, but in order to converge just $t=100$ iterations are needed. In Table \ref{tab:times} is reported the computational time for the convergence of this algorithm. The computations were executed on a processor Intel(R) Core(TM) i7-3720QM and processor speed of 2.60 GHz with 8GB of RAM.

\begin{table}[]
\centering
\renewcommand{\arraystretch}{1.33}
\begin{tabular}{|l|l|l|}
\hline
\textbf{Algorithm} & \textbf{\# of Iterations} & \textbf{Time {[}ms{]}} \\ \hline
HITS & 120 & 56 \\ \hline
Simple PageRank & 185 & 180 \\ \hline
PageRank with Teleportation & 53 & 164 \\ \hline
\end{tabular}
\bigskip
\vspace{-0.2cm}\caption{Number of iterations and time for convergence of the proposed ranking algorithms with $\epsilon=10^{-8}$.}
\label{tab:times}
\end{table}

Finally the predictive power of HITS based on authorities, defined as the percentage of times the higher ranked player will win, is reported in Table \ref{tab:predictive}. For this calculations have been considered data up to the end of August 2017 and the predictive power has been computed as follows. First the new matches played between September 2017 and the end of the 2017 tennis season, for a total number of $431$ matches (those are independent data since are not considered in the training dataset); then all the matches played in 2017 for a total number of $2633$ matches have been considered. As \textit{Modified ATP} is meant that the player who has obtained more ATP points in his career will win. As regards the smaller dataset, \textit{HITS} behave well and similar to the \textit{Modified ATP} system, while for the largest dataset the performances deteriorate.

\begin{figure}[]
\centering
\hspace{0cm}\includegraphics[width=\linewidth]{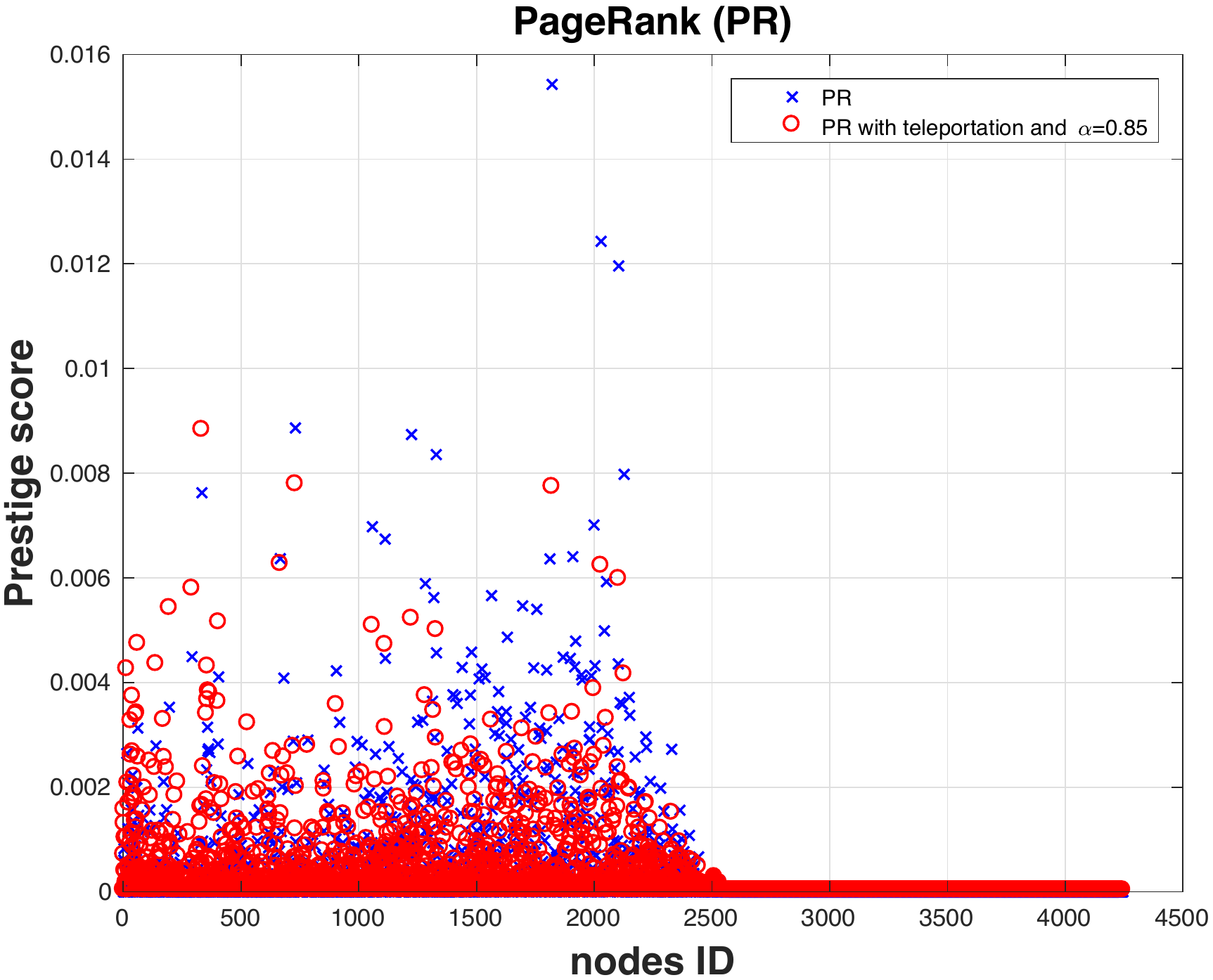}
\vspace{-0.7cm}\caption{Prestige scores of PageRank algorithms.}
\label{fig:PR}
\end{figure}

\begin{table*}[]
\centering
\renewcommand{\arraystretch}{1.33}
\begin{tabular}{c|c|c|c|c|c|c|c|c|}
\cline{2-9}
                                          & \multicolumn{4}{c|}{\textbf{New Data: from 01/09/17 to 30/11/17}}                                       & \multicolumn{4}{c|}{\textbf{New Data: all 2017}}                                                        \\ \hline
\multicolumn{1}{|c|}{\# of Matches}       & \multicolumn{4}{c|}{431}                                                                                & \multicolumn{4}{c|}{2633}                                                                               \\ \hline
                                          & \makecell{\textbf{Modified}\\ \textbf{ATP}} & \makecell{\textbf{Authorities} \\ \textbf{HITS}} & \makecell{\textbf{Simple}\\ \textbf{PR}} & \makecell{\textbf{PR with}\\ \textbf{teleportation}} & \makecell{\textbf{Modified}\\ \textbf{ATP}} & \makecell{\textbf{Authorities} \\ \textbf{HITS}} & \makecell{\textbf{Simple}\\ \textbf{PR}} & \makecell{\textbf{PR with}\\ \textbf{teleportation}} \\ \hline
\multicolumn{1}{|l|}{Right prediction \%} & 59.53\%               & 59.53\%                   & 60.70\%            & 58.84\%                        & 60.92\%               & 60.08\%                   & 60.46\%            & 60.27\%                        \\ \hline
\end{tabular}
\bigskip
\vspace{-0.2cm}\caption{Predictive power of the proposed ranking algorithms with data up to August 2017 and on two different test sets.}
\label{tab:predictive}
\end{table*}
The players prestige scores obtained through PageRank algorithms are plotted in Figure \ref{fig:PR}, where, similarly as before, we can see that the players who only lose matches have the same minimum value.\\
The Top-20 tennis players identified by those algorithm are reported in fourth and fifth columns of Table \ref{tab:rankings}. Without teleportation the podium remains the same as in the authorities of HITS algorithm, then there are many differences. With teleportation we are able to break the loops leading to the biggest authorities and achieve a fairer result.\\
Moreover those results are quite robust and they do not vary much by setting another value to $\alpha$.\\
The fifth column of this table should confirm the goodness of the proposed model being the results very similar to the ones reported in \cite{radicchi}. Actually this table can update the one shown in the mentioned paper where were used data up to 2010 and the resulting top-players were: \textit{Jimmy Connors, Ivan Lendl, John McEnroe, Guillermo Vilas, Andre Agassi, Stefan Edberg, Roger Federer, Pete Sampras, Ilie Nastase, Bjorn Borg, Boris Becker, Arthur Ashe, Brian Gottfried, Stan Smith, Manuel Orantes, Michael Chang, Roscoe Tanner, Eddie Dibbs, Harold Solomon} and \textit{Tom Okker}. Comparing those results with the fifth column of Table \ref{tab:rankings} we can appreciate how the players who are still in activity (\textit{Roger Federer, Rafael Nadal, Novak Djokovic, Andy Murray} and \textit{David Ferrer}) have gained some positions in the overall ranking. It should be stressed that those results are inherently biased toward already retired players, since still active players did not played all the matches of their career; this bias, however, could be removed, for example considering only matches played the same year, as done in \cite{radicchi}. For example, last year (2017) ranking comparisons are reported in the last three columns of Table \ref{tab:rankings} where we see that authorities and PageRank involve mostly the same players in slightly different orders, also with respect to the Official method.

\begin{figure}[]
\centering
\hspace{0cm}\includegraphics[width=\linewidth]{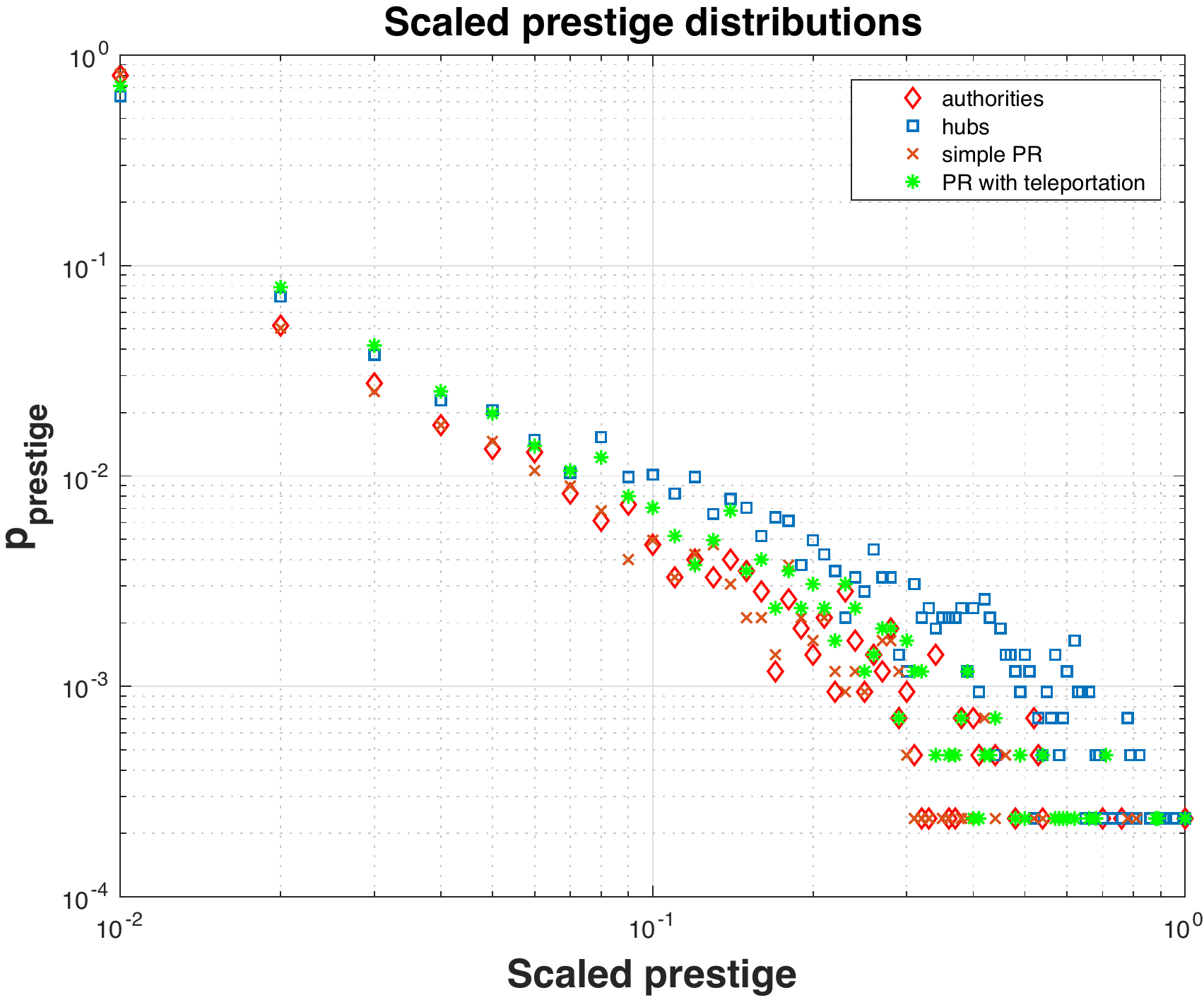}
\vspace{-0.7cm}\caption{Scaled version of prestige scores distributions for the proposed algorithms in log-log plot.}
\label{fig:prestige}
\end{figure}

Moreover, in Figure \ref{fig:prestige} probability distributions of prestige scores obtained through the proposed algorithms are shown. Notice that both $\sum_{i=1}^N \mathrm{prestige}_i=1$ and $\sum_{i=1}^N \mathrm{P[prestige}_i\mathrm{]=1}$, but in this plot the prestige values are reported in a common scale in order to compare the behaviors.

We can see that all the discussed ranking methods behave in a similar way: they have a lot of occurrences of small prestige nodes and a decreasing number of even more prestigious players, where the concept of \textit{prestige} is defined by the specific algorithm. However the probability of highly prestigious players is not negligible since the behaviors follow heavy-tailed distributions. 

The computational demand of the proposed algorithms using $\epsilon=10^{-8}$ is reported in Table \ref{tab:times} and we ascertain that there is no need of speeding-up techniques for our purpose since \textit{N} is not too large.

The predictive power of those algorithms is shown in Table \ref{tab:predictive}. In our analysis PageRank and \textit{Modified ATP} ranking behave similarly and larger test sets are needed to investigate better the results. As order of magnitude the obtained results are consistent with the ones shown in \cite{dingle} but a more robust \textit{ATP} estimator has been achieved by considering all the points gained by a player (called it \textit{Modified ATP}) and not the ATP ranking at the exact time of the match, which is done by the Official ATP estimator, but it has already been proven to achieve worst estimates than e.g. PageRank \cite{dingle}.

\begin{figure}[]
\centering
\hspace{0cm}\includegraphics[width=\linewidth]{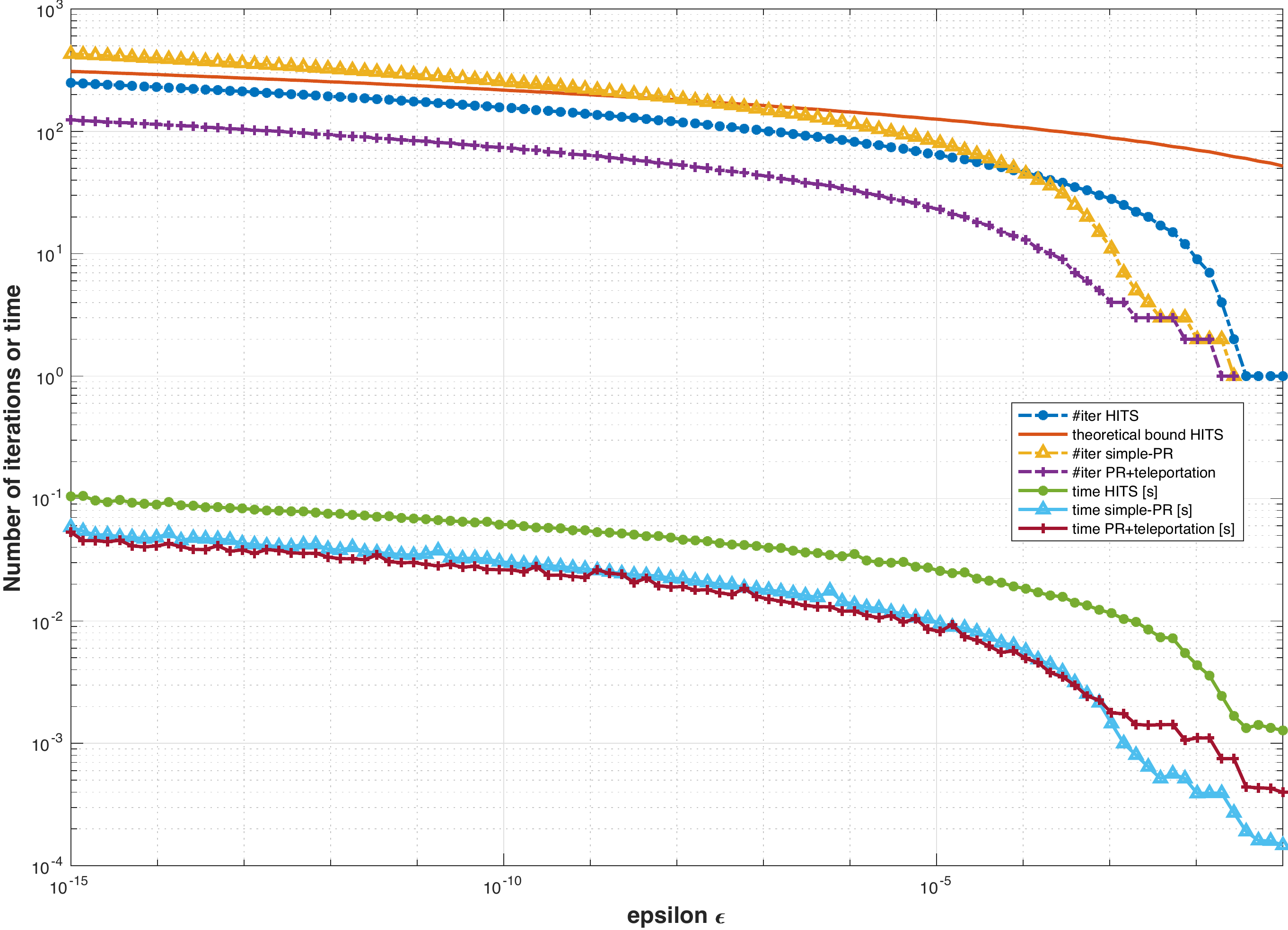}
\vspace{-0.7cm}\caption{Number of iterations and amounts of time needed in order to run the proposed algorithms.}
\label{fig:times}
\end{figure}

Finally, in Figure \ref{fig:times} is shown a comparison of the complexity of the proposed algorithms by varying the convergence parameter $\epsilon$, both in terms of number of iterations and elapsed time. We can appreciate that even though the number of iterations needed by PageRank with teleportation is smaller than the others, the update step is more complex thus resulting in a computational time similar to the simple PageRank. Also, HITS algorithm performs worst than the others in terms of time needed and we can notice that the theoretical bound on its number of iterations is quite strict for small values of $\epsilon$.

\subsection{Link Prediction}
In this section we are going to briefly investigate the players who are likely to play against in future, given that they have never played against before. This is actually a problem of link prediction and we can also consider the undirect and unweighted network's representation because we are looking for predictions at link level, not at the specific outcome of the match. The idea is that \textit{similar} nodes are likely to build a link between them. Firstly, as similarity metric it is used the idea of \textit{Common Neighbors} (CN) defined as: $S_{CN}(i,j)=|\mathcal{N}_i \cap \mathcal{N}_j|$ where $\mathcal{N}_x$ is the set of neighbors of node $x$. Actually, for undirect networks a simple expression holds: $\mathbf{S}_{CN}=\mathbf{A}^2$. Moreover we need to restrict our attention to active players, thus they could effectively play a match in future. 

Applying all those considerations above we found that the six most likely matches to be drawn are: \textit{Victor Troicki - Ivo Karlovic}, \textit{Rafael Nadal - Yen Hsun Lu}, \textit{Teymuraz Gabashvili - Gael Monfils}, \textit{Marin Cilic - Dmitry Tursunov}, \textit{Nicolas Mahut - Marcos Baghdatis} and \textit{Fabio Fognini - Nicolas Mahut}.

Applying the general framework of link prediction in complex networks, i.e. removing 10\% of the link for each network and performing 100 iterations of the link prediction algorithm, it is possible to assign ikelihood scores to all the non-observed links in the reduced network. In order to evaluate the performance, the links are ranked by likelihood scores and the precision is computed as the percentage of removed links among the top-r in the ranking, where r is the total number of links removed.
The link prediction algorithms used here are CN, Resource Allocation (RA) and Structural Perturbation Method (SPM) \cite{SPM}; the results are reported in Table \ref{tab:linkPred}, where we see that SPM is the best method among them.

\begin{table}[]
\centering
\caption{Mean precision of link prediction algorithms on the two undirect networks.}
\label{tab:linkPred}
\begin{tabular}{c|lll}
\multicolumn{1}{l|}{Network \#} & CN & RA & SPM \\ \hline
3                               &  0.304  &  0.295  & 0.316    \\
4                               &  0.146   &    0.152 &   0.204
\end{tabular}
\end{table}

\subsection{Communities Detection}

It can be interesting to partition the graph in $k$ disjoint groups, \textit{communities}, for example through \textit{spectral clustering} technique defined in \cite{shi}. A community is a group of nodes who have a higher likelihood of connecting to each other than to nodes from other communities. Intuitively, one should expect that $k$ communities will appear, each containing players of the same era. However it is an interesting problem how to find the best partition such that minimizes the connections among the $k$ groups.

For simplicity let's consider the case of $k=2$, any other choice is a straightforward extension.  Consider the normalized Laplacian matrix $\displaystyle \breve{\mathbf{L}}=\mathbf{I-D^{-\frac{1}{2}}A D^{-\frac{1}{2}}}$, where $\mathbf{D}=diag(\mathbf{d})$ and $\mathbf{d=A1}$. The normalization makes the Laplacian matrix more \textit{stable} in the sense that the produced eigenvectors are less noisy. Then find the second largest eigenvalue $\lambda_{N-1}$ and its eigenvector $\mathbf{v}_{N-1}$ respectively called \textit{algebraic connectivity} and \textit{Fiedler vector} from \cite{fiedler}: hence in order to find the two communities one can simply look at the sign of the Fiedler vector and assign indices corresponding to positive values to one community and vice versa (otherwise one can resort to more sophisticate clustering algorithms). 

\begin{figure}[]
\centering
\hspace{0cm}\includegraphics[width=\linewidth]{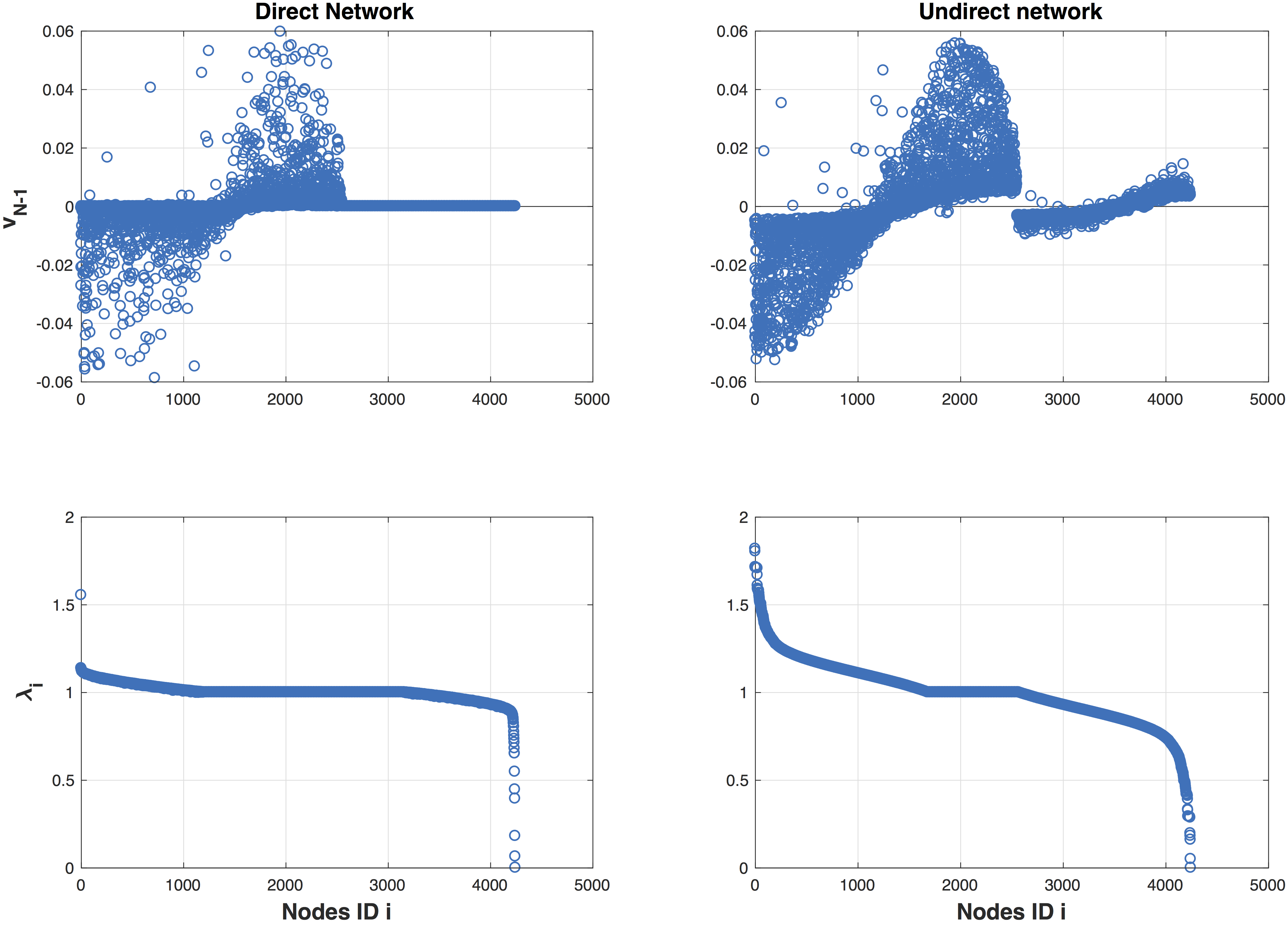}
\vspace{-0.7cm}\caption{Fiedler vector $\mathbf{v_{N-1}}$ and eigenvalues $\lambda_i$ for direct and undirect network.}
\label{fig:fiedler}
\end{figure}

Figure \ref{fig:fiedler} reports the Fiedler vectors $\mathbf{v_{N-1}}$ and all the eigenvalues $\lambda_i$ for both the direct and undirect representations. First of all we can confirm that $\lambda_N=0$ and $\lambda_{1}<2$ as we expect from theoretical analysis. Then we can notice that only two or three eigenvalues can be considered small and the eigengap between them is still quite large; hence a partitioning in two or three communities is \textit{a posteriori} sensible. Moreover defining a conductance measure $h_G=\min_A  \frac{\mathrm{cut}(A,A^C)}{\min(\mathrm{assoc}(A),\mathrm{assoc}(A^C))}$, the Cheegar's inequality $\frac{1}{2}\lambda_{N-1}\leq h_G \leq \sqrt{2 \lambda_{N-1}}$ helps in measuring the quality of spectral clustering: more specifically a low value of $h_G$ means that the partitioning is good. Were found $0.0321\leq h_g \leq 0.3585$, where the upper bound is not very small, thus the partition will not be very accurate, because we need to divide the careers of many peer players, thus many links will exist between the communities.

From the sign of the Fiedler vector we can see that the previous intuition was correct and we can identify 1988 as the year of transition (i.e. around player ID 1400). That year is not at all the half of the considered period, which goes from 1968 to 2017; it indicates, instead, the year of a seminal moment in ATP history, because in 1988 ''The Parking Lot Press Conference'' \cite{parking_lot} took place, which states the beginning of the ATP Tour era. From there onward tennis match schedules are similar to what we are used to nowadays while before the tennis circuit was very different.

\section{Conclusions and Future Work}

In this study we have shown how it is possible to map the ATP single tennis matches into different graph representations; then we evaluated some metrics typical of those networks and we compared the results with the existing literature compensating for the lack of structural analysis of such networks.

We have performed a joint analysis of few different ranking techniques and we have evaluated them showing analogies and differences, also comparing and extending the results already present in literature. We have shown that \textit{Jimmy Connors} is still the best player in tennis history up to 2017 according to the PageRank with teleportation algorithm, but actually \textit{Roger Federer} is approaching the top position, indeed it is at about the same value of \textit{Ivan Lendl}. If he will succeed in winning again in 2018 most of the more important tournaments, he will be definitively the best of all the times by the end of 2018.

An interesting aspect of the proposed ranking systems is that they do not require any arbitrary introduction of external criteria for the evaluation of the quality of players and tournaments. Players' \textit{prestige} is in fact self-determined by the network structure. The proposals achieve also similar predictive power to the modified ATP ranking and defeat the official one.

Those considerations on predictive power should be reinforced in the near future by choosing an enlarged test set. In future, for example, we would like to include in the statistic the new matches played in 2018, in order to have independent data, and also include and evaluate other modifications to PageRank algorithm.

Moreover we have briefly discussed about link prediction methods, and an extensive validation could be performed applying also other algorithms.

Then we have seen an interesting and powerful application of spectral clustering for graph partitioning and we have recognized a promising result. We can further investigate how the partitions will change by increasing the number of cluster or by using a different communities detection algorithm.

%\IEEEtriggeratref{11}
\bibliography{biblio}

\end{document}